\documentclass[iop,revtex4]{emulateapj}
\usepackage{natbib}
\usepackage{graphicx}
\usepackage{threeparttable,lscape}
\usepackage{rotating}
\usepackage{hyperref}
\usepackage{color}
\usepackage{xcolor}
\usepackage{adjustbox}      

\definecolor{yaleblue}{rgb}{0.1,0.3,0.9}
\definecolor{ultramarine}{rgb}{0, 0, 150}
\hypersetup{colorlinks=true, linkcolor=black, urlcolor=blue, citecolor=yaleblue}

\newcommand{\Msun}{M$_\odot$~}

\shorttitle{SNe~II: spectroscopic and photomeric correlations}
\shortauthors{Guti\'errez et al.}

\begin{document}

\title{Type II supernova spectral diversity II: Spectroscopic and photometric
correlations\footnote{T\lowercase{his 
paper includes data gathered with with the 6.5 m} M\lowercase{agellan} 
T\lowercase{elescopes located at} L\lowercase{as} C\lowercase{ampanas} O\lowercase{bservatory,} 
C\lowercase{hile; and the} G\lowercase{emini} O\lowercase{bservatory,} C\lowercase{erro} P\lowercase{achon,}
C\lowercase{hile} (G\lowercase{emini} P\lowercase{rogram} GS-2008B-Q-56). B\lowercase{ased on
observations collected at the} E\lowercase{uropean} O\lowercase{rganisation for} A\lowercase{stronomical}
R\lowercase{esearch in the} S\lowercase{outhern} H\lowercase{emisphere,} C\lowercase{hile} 
(ESO P\lowercase{rogrammes} 076.A-0156, 078.D-0048, 080.A-0516, \lowercase{and} 082.A-0526).}.}

\author{Claudia P. Guti\'errez\altaffilmark{1,2,3,4}
Joseph P. Anderson\altaffilmark{3},
Mario Hamuy\altaffilmark{2,1},
Santiago Gonz\'alez-Gaitan\altaffilmark{1,5,6},
Lluis Galbany\altaffilmark{7},
Luc Dessart\altaffilmark{8},
Maximilian D. Stritzinger\altaffilmark{9},
Mark M. Phillips\altaffilmark{10},
Nidia Morrell\altaffilmark{10},
Gast\'on Folatelli\altaffilmark{11}}

\altaffiltext{1}{Millennium Institute of Astrophysics, Casilla 36-D, Santiago, Chile }
\altaffiltext{2}{Departamento de Astronom\'ia, Universidad de Chile, Casilla 36-D, 
Santiago, Chile}
\altaffiltext{3}{European Southern Observatory, Alonso de C\'ordova 3107, Casilla 19, Santiago, Chile}
\altaffiltext{4}{Department of Physics and Astronomy, University of Southampton, Southampton, SO17 1BJ, UK}
\email{C.P.Gutierrez-Avendano@soton.ac.uk}
\altaffiltext{5}{Center for Mathematical Modelling, University of Chile, Beauchef 851, Santiago, Chile}
\altaffiltext{6}{CENTRA, Instituto Superior T\'ecnico - Universidade de Lisboa, Portugal}
\altaffiltext{7}{PITT PACC, Department of Physics and Astronomy, University of Pittsburgh, Pittsburgh, PA 15260, USA}
\altaffiltext{8}{Unidad Mixta Internacional Franco-Chilena de Astronom\'ia (CNRS UMI 3386), Departamento de Astronom\'ia, 
Universidad de Chile, Camino El Observatorio 1515, Las Condes, Santiago, Chile}
\altaffiltext{9}{Department of Physics and Astronomy, Aarhus University, Ny Munkegade 120, DK-8000 Aarhus C, Denmark}
\altaffiltext{10}{Carnegie Observatories, Las Campanas Observatory, Casilla 601, La Serena, Chile}
\altaffiltext{11}{Facultad de Ciencias Astron\'omicas y Geof\'isicas, Universidad Nacional de La Plata, Instituto de 
Astrof\'isica de La Plata (IALP), CONICET, Paseo del Bosque S\/N, B1900FWA La Plata, Argentina} 

\begin{abstract}
We present an analysis of observed trends and correlations between a large range of spectral and photometric
parameters of more than 100 type II supernovae (SNe~II), during the photospheric phase. We define a common
epoch for all SNe of 50 days post-explosion where the majority of the sample is likely to be under similar physical
conditions. Several correlation matrices are produced to search for interesting trends between more than 30 distinct
light-curve and spectral properties that characterize the diversity of SNe~II. Overall, SNe with higher expansion
velocities are brighter, have more rapidly declining light-curves, shorter plateau durations, and higher $^{56}$Ni
masses. Using a larger sample than previous studies, we argue that `$Pd$' - the plateau duration from the transition of 
the initial to `plateau' decline rates to the
end of the `plateau' - is a better indicator of the hydrogen envelope mass than the traditionally used optically 
thick phase duration ($OPTd$: explosion epoch to end of plateau). This argument is supported by the fact that $Pd$ also correlates
with s$_3$, the light-curve decline rate at late times: lower $Pd$ values correlate with larger s$_3$ decline rates.
Large s$_3$ decline rates are likely related to lower envelope masses that enables gamma-ray escape.
We also find a significant anticorrelation between $Pd$ and s$_2$ (the plateau decline rate),
confirming the long standing hypothesis that faster declining SNe~II (SNe~IIL) are the result of explosions with
lower hydrogen envelope masses and therefore have shorter $Pd$ values.
\end{abstract}


\section{Introduction}

It is commonly accepted that Core-Collapse Supernovae (CC-SNe) are produced by the explosion of massive 
($>8$ M$_\odot$) stars. CC-SNe display a
wide spectral and photometric variety, leading to the basis of their spectral classification. 
First order CC-SN classification is based on the presence or absence of hydrogen 
within SN spectra. SNe where hydrogen is clearly visible are called SNe~II, while those without these 
features correspond to SNe Ib/c \citep{Minkowski41,Filippenko97}. \\
\indent Initially, SNe~II were classified according to the shape of the light curve: SNe with a faster decline 
rate are called SNe~IIL, while SNe with almost constant luminosity for several months were called SNe~IIP 
\citep{Barbon79}. However, years later, two new classes of SNe~II emerged:
SNe~IIn and SNe~IIb. SNe~IIn show narrow emission lines in their spectra, possibly due to steady interaction with a
circumstellar medium (CSM; \citealt{Schlegel90}), while SNe~IIb are thought to be transitional events between SNe~II 
and SNe~Ib \citep{Filippenko93}. 
The overall properties of SNe~IIn and SNe~IIb are sufficiently distinct from `normal' SNe~II,
that we do not include them for study, and they are no longer discussed in this paper.\\
\indent  With ever increasing numbers of SNe, new sub-classes have appeared. 
\citet{Blanco87,Menzies87,Hamuy88,Phillips88} and \citet{Suntzeff88} presented analysis
of SN~1987A, an object that exhibited typical characteristics of the SN~II spectra, but a peculiar light
curve. With this SN the 87A-like objects were introduced. Examples of these SNe can be found in
\citet{Pastorello05}, \citet{Pastorello12}, and \citet{Taddia13}\footnote{As the SN~87A-like objects have different light-curve 
properties than `normal' SNe~II, we also exclude them from our analysis.}. Later, 
\citet{Pastorello04} and more recently \citet{Spiro14} studied the properties of low 
luminosity SNe~II, which additionally have narrow spectral lines (indicating low expansion velocities). 
On the other hand, \citet{Inserra13} analyzed a group of luminous SNe~II. Lately, intermediate luminosity 
SNe have been also studied, supporting the wide diversity in SNe~II \citep[e.g.][]{Roy11,Takats14}.\\
\indent Red Super-Giant (RSG) stars with zero-age main-sequence mass $\geq$ 8 M$_\odot$~ have generally been 
assumed as the progenitors of SNe~II, with hydrodynamical modelling supporting this hypothesis
\citep{Chevalier76}. In recent years, a significant number of direct identifications
of the progenitor stars of nearby SNe~IIP \citep[e.g.][]{VanDyk03, Smartt04, Smartt09, Maund05, Smartt15}
suggest that RSG stars with masses of 8 - 18 M$_\odot$~  are their progenitors, supporting 
initial assumptions. There is little observational constraint on the progenitor mass range of SNe~IIL because
only two direct identifications have been obtained (\citealt{Elias-Rosa10,Elias-Rosa11}, but see 
\citealt{Maund15}), however these do provide some evidence in favor of higher mass progenitors.
Nevertheless, a recent analysis done by \citet{Valenti16} with the light curves and spectra of 16 SNe~II 
did not find any evidence for progenitor mass differences between SNe of different decline rates. \\
\indent While direct detections of progenitors have constrained a relatively narrow mass range for SNe~II,
the same SNe show significant differences in their final explosive displays (e.g. SN~2004et, a normal SNe~II, and
SN~2008bk, a low luminosity event). It must therefore be that differences in stellar evolutionary 
processes leave the progenitors in different final states (e.g. the extent of the hydrogen envelope,
the progenitor radius at explosion, the CSM) or explode with e.g. different energies, in order to produce the 
diversity we observe.\\
\indent Theoretical studies have suggested that progenitors that explode with smaller hydrogen
envelope masses produce faster declining light curves (SNe~IIL), together with shorter or non-existent 
`plateaus' \citep[e.g.][]{Litvinova83,Bartunov92,Popov93,Morozova15,Moriya15}. An alternative study presented by  \citet{Kasen09} shows 
that a change in the explosion energy leads to a range of luminosities, velocities, and 
light curve durations. That is to say, higher explosion energies result in brighter events
with higher expansion velocities and shorter plateaus.
They also found that an increasing synthesised $^{56}$Ni  mass extends the
length of the plateau (see also \citealt{Bersten13T}).
Meanwhile, \citet{Dessart13} using radiative-transfer models explored
the properties of SNe~II changing the physical parameters of the progenitor and/or the explosion (e.g. metallicity, explosion energy,
radius). They found that the radius has an influence on the temperature/ionisation/color evolution (more compact
objects cool and recombine faster) and in the plateau
brightness, while a variation in the explosion energy leads to a variation of the plateau 
brightness and the plateau duration, consistent with \citet{Kasen09}. \\
\indent To quantify the spectral and photometric diversity, a number of statistical studies of SNe~II have been 
published. \citet{Patat94} characterized the properties of 57 SNe~II using the maximum
$B$-band magnitude, the color at maximum and the ratio of emission to absorption ($e/a$) in H$_{\alpha}$.
They showed that faster declining events are more luminous, have shallower P-Cygni profiles 
and are bluer than SNe~IIP. The majority of more recent studies have focused on SNe~IIP.
\citet{Hamuy02} analyzed 17 SNe~IIP and found that SNe with brighter plateaus have
higher expansion velocities (also seen in the models of \citealt{Bersten13T}.
\citet{Hamuy03} concluded that more massive SN~IIP progenitors produce
more energetic explosions and in turn produce more nickel. Similar results were found by \citet{Pastorello03} and 
more recently by \citet{Faran14a}.  The only exception to these works about SNe~IIP was published by
\citet{Faran14b}, who analyzed a sample of SNe~IIL. They found that faster declining SNe~II (SNe~IIL)
are brighter than slower declining events (SNe~IIP), confirming previous results.\\
\indent \citet{Gutierrez14} and \citet{Anderson14a} 
using a large sample of SNe~II, analyzed the dominant line in SNe~II, the H$_{\alpha}$ P-Cygni profile.
\citet{Gutierrez14} using a sample of 52 SNe~II (a sub-sample of that which we present here) showed that SNe with
smaller values of $a/e$ (the inverse of the ratio previously discussed by \citealt{Patat94})
are brighter and have faster declining light curves. They concluded that these relationships and the diversity of $a/e$
can be understood in terms of a varying hydrogen envelope mass at explosion epoch, together with the possibility of an
influence of circumstellar interaction. 
Meanwhile, \citet{Anderson14a} analyzed the blueshifted offset in the emission peaks of H$_{\alpha}$ 
of 95 SNe~II. Through comparison to spectral modelling \citep{Dessart05,Dessart13a}, they argue that this behaviour
is a natural consequence of the distinct density profiles found in SN ejecta.\\
\indent Using a sample of 117 SNe~II, \citet{Anderson14} (hereafter A14)  studied the $V$-band light curve diversity of these objects. 
They found that SNe~II with shorter plateau duration ($Pd$) exhibit faster decline rates (s$_2$ in their nomenclature). 
They concluded that the envelope mass at the epoch of explosion is the  dominant physical parameter that 
explains this observed diversity. Similar results were found by \citet{Sanders14},
\citet{Valenti16} and \citet{Galbany16}.  They also found that SNe~IIP and SNe~IIL show a continuum in their 
photometric properties and it is not suitable to isolate them in two distinct classes or types.\\
\indent In addition to these results, A14 found relatively high 
radioactive decline rates ($s_3$) for a significant number of SNe. 
In $^{56}$Ni powered light curves at late times, full gamma-ray and positron trapping yields a decline 
rate s$_3$ of 0.98mag per 100 days. Higher decline rates than this value therefore suggest less 
efficient trapping of gamma-ray emission (or much greater explosion energies),
suggesting lower mass ejecta for these SNe~II.\\
\indent The previous discussion shows how numerous relations between observed photometric and spectral 
parameters have been used to understand the SN~II phenomenon. However, there are many additional parameters
that have not been included in this discussion to date. Inclusion of additional parameters can aid in furthering 
our understanding of the underlying physics of SNe~II.
This motivates our current work where we study a sample of almost 1000
optical-wavelength spectra of $>100$ SNe~II. To that aim, we have divided the analysis into two papers.
In Gutierrez et al. (2017) (hereafter Paper I) we present the full description of the observations, data reduction techniques, and the spectral 
properties. We also discuss the spectral matching technique to estimate the explosion epochs, the analysis of
the spectral line evolution and the nature of the extra absorption component on the blue side of H$_{\alpha}$. \\
\indent Here, in this paper II we analyse the correlations between different spectral parameters
defined to explore the diversity of SNe~II, together with their correlation with previously defined photometric
measurements. Expansion velocities, pseudo-equivalent widths (pEWs), the ratio of 
absorption to emission ($a/e$) of the H$_{\alpha}$ P-Cygni profile, and velocity decline rates
are used to search for correlations with photometric parameters and between other spectral properties.
We analyze spectral correlations and determine the most important properties to compare them with 
the photometric parameters. Our overall aim is to search for trends between different measured parameters, 
and then attempt to link these to the underlying physical properties of SN II progenitors.\\
\indent The paper is organized as follows. Section~\ref{data}  briefly describes the data employed 
for this analysis. In Section ~\ref{measurements} we describe our measurement techniques.
An overall current physical understanding of our different observed parameters is presented in Section~\ref{param}. 
The full analysis is presented in Section~\ref{results}. 
We discuss our results in Section~\ref{discussion} and present our conclusions in Section~\ref{conclusions}. \\

\section{Data}
\label{data}

The data used in this analysis were published in A14 and Paper I.
The details of the spectroscopic and photometric observations and reductions can be found in 
the mentioned studies. On average we have 7 spectra per SN, which are analysed together with their 
$V$-band light-curves. Details of these SNe are available in A14, \citet{Anderson14a},
\citet{Gutierrez14}, \citet{Galbany16} and Paper I.\\
\indent A small number of SNe presented in Paper I are excluded from this work because they 
have insufficient spectral and/or photometric data to be useful
(SNe~1988A,~1990E,~1992ad,~1992am, 1993A, 1999eg, 2002ew, 2003dq, 2004dy, 2005dw, 
2005es, 2005K, 2005me, 2006bc, 2007Z, 2008F, 2009W).
 
\section{Measurements}
\label{measurements}

The evolution of SNe~II can be studied according to both spectral and photometric behaviour. 
At early phases the spectra exhibit the Balmer lines (H$_{\alpha}$, H$_{\beta}$, H$_{\gamma}$,
H$_{\delta}$), and \ion{He}{1} $\lambda5876$ \AA. With time, the iron group lines start to appear and to 
dominate the region between 4000 and 6000 \AA. The \ion{Ca}{2} triplet, \ion{Na}{1} D, and
\ion{O}{1} also emerge. The light curve at the beginning shows a fast rise to peak brightness, followed by 
a slight decline, which is powered by the release of shock deposited energy. 
Around $\sim30$ days post-explosion a plateau arises from 
the fact that the expansion of the ejecta at the photosphere compensates for the drop in optical depth.
When the photospheric phase ends (around 80-120 days post explosion, A14),
the transition to the nebular phase starts and the brightness drops. Once this happens,
the radioactive tail phase starts. This phase is powered by the 
radioactive decay of $^{56}$Co to $^{56}$Fe. Later than $\sim200$ days, the spectra are dominated 
by forbidden lines, which are formed in the inner part of the ejecta. 
Much diversity is observed both in spectra and photometry, 
which suggests differences in the properties of the progenitor star and the explosion.\\
\indent To study the diversity within SNe~II we use the spectral and photometric parameters defined in
\citet{Gutierrez14} and A14. We also define a number of additional parameters below. 
These measurements are chosen to enable a full characterisation of the diversity of SN~II $V$-band
light curves and optical spectra.

\subsection{Spectral measurements}

Before proceeding with our spectral analysis, below we summarise the parameters we use, as
defined in Paper I:
\begin{itemize}
\item $v$: corresponds to the expansion  velocity. It is measured from the minimum flux of the 
absorption component of P-Cygni line profile. In this analysis we measure this parameter for eleven 
features in the photospheric phase: H$_{\alpha}$, H$_{\beta}$, \ion{Fe}{2} $\lambda$4924, \ion{Fe}{2} $\lambda$5018,
\ion{Fe}{2} $\lambda$5169, \ion{Sc}{2}/\ion{Fe}{2} $\lambda$5531, \ion{Sc}{2} multiplet $\lambda$5663, 
\ion{Na}{1} D, \ion{Ba}{2} $\lambda$6142, \ion{Sc}{2} $\lambda$6247, and  \ion{O}{1} $\lambda$7774.
In the case of H$_{\alpha}$, the velocity was also derived using the full-width-at-half-maximum (FWHM) of the emission component. 
\item $\Delta v$(H$_{\beta})$: defined as the rate of change of the expansion velocity of the H$_{\beta}$ feature.
This parameter was measured at 5 distinct intervals (see Paper I), however here we only use
the interval $50\leq t \leq 80$ days,  as this shows the highest correlation with other parameters.
\item $\Delta vel$: defined as the velocity difference between H$_{\alpha}$ and Fe II $\lambda$5018, and Na I D and 
Fe II $\lambda$5018. 
\item pEW: corresponds to the absorption/emission strength of a particular line. Here, we measure the 
absolute value of pEW for the same features mentioned above. 
\item $a/e$: defined as the flux ratio of the absorption to emission component of H$_{\alpha}$ P-Cygni profile.
This ratio is the inverse of that presented by \citet{Patat94}. We propose $a/e$ as this deals better with
weak absorption values that are shown by a number of SNe~II in our sample.
\end{itemize}

\indent  While measurements were performed in all epochs at which we obtained spectra, we choose to define common
epochs between SNe at 30, 50 and 80 days post explosion. An interpolation and extrapolation is used to obtain parameter
values at these epochs.
The values obtained by the interpolation are used when two available spectra are present $\pm15$ days around
the common epoch, while the values from the extrapolation are used at $\pm10$ days.  These intervals were chosen 
as they increase the strength of observed correlations. Using bigger intervals deteriorates the correlations
because the polynomial does not produce reliable results in some cases (particularly for the pEW).
At $\pm15$ and $\pm10$ days for interpolation and 
extrapolation, respectively, the results do not show a significant change compared to those obtained using a 
smaller interval. Hence, our choice of intervals is justified. To estimate the velocity at a common epoch, we do an
interpolation/extrapolation using a power law fit. For the pEW we use a low order (first or second) polynomial fit.
Power law fits were found to produce satisfactory results in the case of velocity measurements, however for 
pEWs we found that low-order polynomials were required. For this parameter we used a low order polynomial and
determined the best fit using the normalized root mean square (rms) of different orders.
The errors of each measurement were obtained with the rms error fit.
In summary, we are able to use spectral parameter values in 88, 84, and 59 SNe at 30, 50 and 80 days, respectively. 

\subsection{Photometric measurements}

Historical separation of SNe~II into distinct classes was based on photometric differences 
in e.g. decline rates and absolute magnitudes. Hence it is essential to include photometric parameters in our 
analysis for a full understanding of observed correlations and their implications for SN~II physics. Here, we use
the $V$-band photometric parameters already defined (and measured) in A14, which we now summarise:
\begin{itemize}
\item $t_{0}$: corresponds to the explosion epoch (see Paper I for more details of their estimation).
\item $t_{tran}$: determined as the transition between the initial decline (s$_1$) and the plateau decline (s$_2$).
\item $t_{end}$: corresponds to the end of the optically thick phase (i.e., the end of the plateau phase).
\item $t_{PT}$: is the mid point of the transition from ‘plateau’ to radioactive tail.
\item $OPTd$: is the duration of the optically thick phase and is equal to $t_{end}-t_{0}$.
\item $Pd$: is the plateau duration, defined between  $t_{tran}$ and $t_{end}$.
\item $M_{max}$: defined as the initial peak in the $V-$band light-curve.
\item $M_{end}$: defined as the absolute $V-$band magnitude measured 30 days before t$_{PT}$.
\item $M_{tail}$: defined as the absolute $V-$band magnitude measured 30 days after t$_{PT}$.
\item $s_1$: defined as the decline rate ($V-$band magnitudes per 100 days) of steeper slope of the light-curve.
\item $s_2$: defined as the decline rate ($V-$band magnitudes per 100 days) of the second, shallower slope in the light curve.
\item $s_3$: defined as the linear decline rate ($V-$band magnitudes per 100 days) of the slope in the
radioactive tail part.
\item $^{56}$Ni mass: corresponds to the mass of radioactive nickel synthesised in the explosion. 
(A14 for exact details of how this was estimated).
\end{itemize}

\begin{figure}
\centering
\includegraphics[width=\columnwidth]{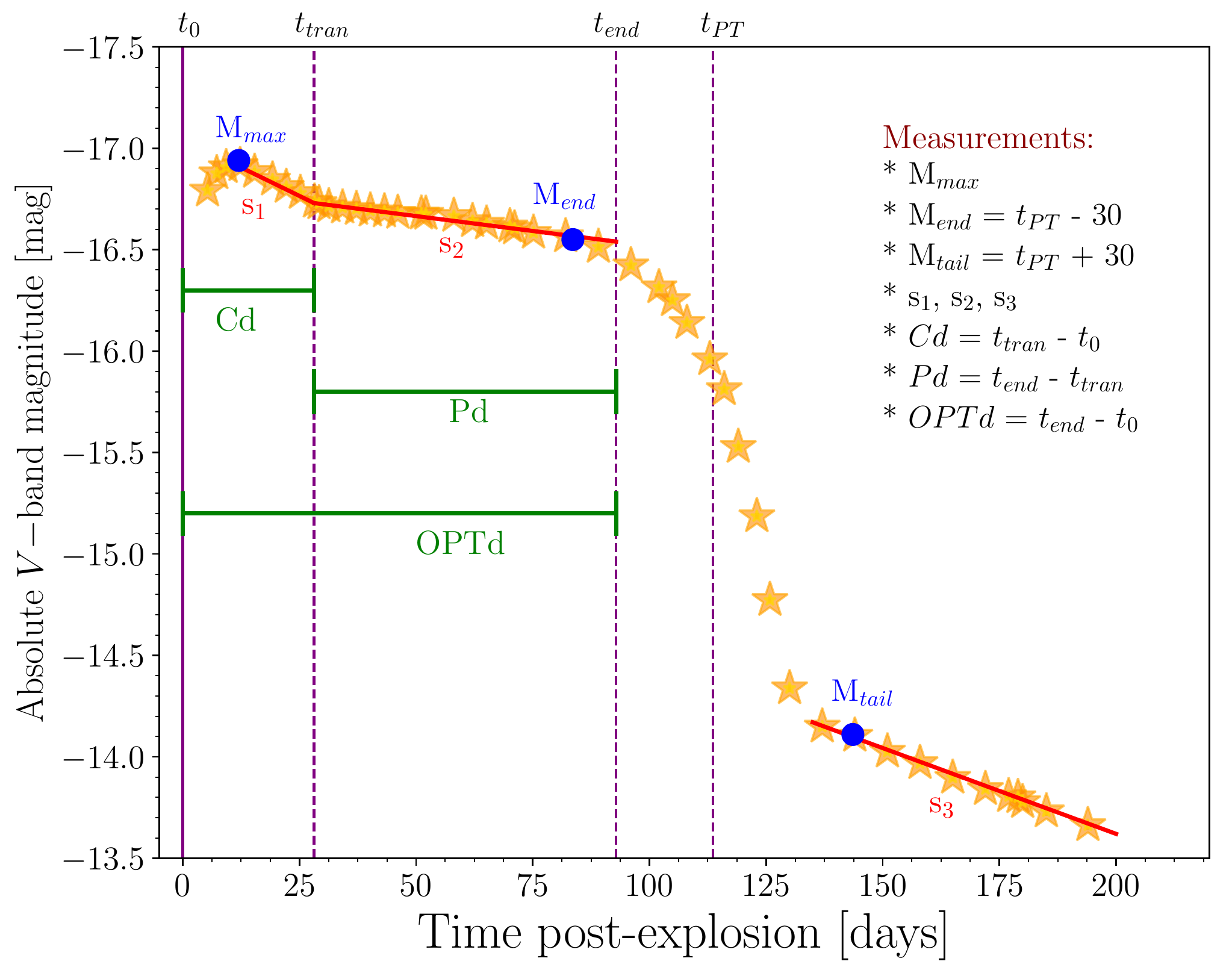}
\caption{Example of the light-curve parameters measured for each SN within the sample in the $V-$band.
Observed absolute magnitude at peak, M$_{max}$, M$_{end}$ and M$_{tail}$ are shown in blue, as applied to the dummy
data points (yellow stars) of a SN~II. The positions of the three measured slopes, s$_1$ , s$_2$, and s$_3$, are
shown in red. The cooling duration ($Cd$), plateau duration ($Pd$) and optically thick phase duration ($OPTd$), are indicated 
in green. Four time epochs are labeled: t$_0$, the explosion epoch; t$_{tran}$, the transition from s$_1$ to s$_2$; 
t$_{end}$, the end of the optically thick phase; and $t_{PT}$, the mid point of the transition from ‘plateau’ to 
radioactive tail.}
\label{lc}
\end{figure}

Initial values for these parameters can be found in Table~5 in A14, however it should be noted that 
in this work some of these parameters have been updated: $t_{tran}$, $OPTd$, $Pd$, M$_{max}$, M$_{end}$, M$_{tail}$,
s$_1$ and s$_2$. In the case of magnitudes it was found that stronger correlations were obtained with other
parameters before any extinction corrections were made. This suggests that
a) in the vast majority of cases host galaxy extinction is relatively small, and b) when we do 
make extinction corrections (using the absorption \ion{Na}{1} D in A14), such corrections are 
not particularly accurate. Therefore, all magnitudes are being used without host galaxy extinction corrections. 
For $t_{tran}$ we used the F-test to decide whether a one or two slope fit was better; A14 used the
BIC criterion. The main difference resides in how the F-test penalises the number of parameters of 
each model (more details in Galbany et al., in prep.). This method increases the number of SNe with $t_{tran}$ available, 
and in turn this increases the number of SNe for which we can define $s_1$ and $Pd$. 
A visual check of those SNe~II showing $t_{trans}$ using both the F-test and the BIC criterion was performed,
and this gives us confidence in the use of the former in this work.
All values used in the current analysis are listed in Table~\ref{photo}.\\
\indent Besides the parameters defined by A14 we include two more parameters:
\begin{itemize}
\item $\Delta (B-V)$: defined as the color gradient. We measure this parameter in three different ranges:
$10\leq t \leq20$d, $10\leq t \leq30$d, and $20\leq t \leq50$d. Color gradients are calculated by
fitting a low-order polynomial to color curves and then taking the color from the fit at each epoch 
and calculating the gradient, $\Delta (B-V)$ by simply subtracting one epoch color from the other and dividing by
the number of days of the interval.
\item $Cd$: corresponds to the cooling phase durations ($Cd$), defined between t$_{0}$ and t$_{tran}$.
\end{itemize}
Figure~\ref{lc} presents an example light curve indicating all the above defined $V-$band parameters.

\section{Observed parameters and their physical implications}
\label{param}

The basic properties of the progenitor stars and explosion that have a significant influence on SN~II
diversity are the explosion energy ($E$), ejecta mass ($M_{\rm ej}$),  pre-supernova radius ($R_0$),  
the $^{56}$Ni mass, and progenitor metallicity (with many of these parameters likely to be directly 
linked to the Zero Age Main Sequence, ZAMS, mass).
Theoretical works \citep[e.g.][]{Young04, Kasen09, Dessart13a} have studied how variations of these 
parameters  influence SN~II light curves and spectra. Specifically, such studies have directly linked
observed parameters such as luminosities, expansion velocities and the duration of the plateau to the
above physical progenitor properties.\\
\indent The most commonly used parameter to link observed SN properties to progenitor characteristics
has been the duration of the plateau. It has been associated to the
hydrogen envelope mass of the progenitor at the moment of the explosion.
Theoretical models \citep[e.g.][]{Litvinova83,Popov93,Dessart10b,Morozova15,Moriya15} 
have shown that the plateau duration is a good 
indicator of the hydrogen envelope mass in the direction that larger envelope masses produce longer 
duration plateaus. This can be understood as the hydrogen recombination wave taking a longer time to 
travel back through the ionised ejecta in SNe with a larger hydrogen envelope. Traditionally, authors 
have referred to the `plateau duration' as the time from explosion to the epoch when each SN starts to
transition to the nebular phase. However, such a definition then includes phases that are powered by 
different physical mechanisms (early-time light curves are powered by the release of shock deposited 
energy, while later phases during the true plateau are powered by hydrogen recombination 
(e.g. \citealt{Grassberg71,Chevalier76,Falk77}). 
In A14 two time durations were defined: $OPTd$, the optically thick phase duration,
and $Pd$ the plateau duration. The former is equivalent to the traditional definition of the plateau 
duration from explosion to the end of the plateau, while the latter is defined from the inflection point
in the s$_1$ and s$_2$ decline rates to the end of
the plateau. The newly defined $Pd$ value should thus more accurately scale with hydrogen envelope mass,
while $OPTd$ includes both effects of changing the envelope mass together with radius differences affecting
the time taking for the light-curve to reach the hydrogen recombination powered s$_2$ decline rate. Later
we provide additional evidence and arguments for this interpretation: overall correlations are stronger 
between $Pd$ and other SN~II measurements (particularly those other parameters linked to the envelope mass)
than $OPTd$.\\
\indent In addition to $Pd$, it was argued in A14 that decline rates during the radioactive 
phase, s$_3$, can also give an indication of the ejecta mass. The expected s$_3$ decline rate is 0.98 mag per 100 days
assuming full trapping of the radioactive emission from $^{56}$Co decay \citep{Woosley89}.\\
\indent The expansion velocity and luminosity of SNe~II are both set by the explosion energy (\citealt{Kasen09} and \citealt{Bersten13T}):
more energetic explosions produce higher photospheric velocities, and in turn, brighter events. 
These results have been showed observationally by \citet{Hamuy02L,Hamuy03}.\\
\indent More recently, \citet{Dessart10}; \citet{Dessart13a} showed that in SNe with small progenitor radii,
the recombination phase starts earlier.
This would imply that the phase between the explosion and $t_{tran}$ (cooling duration phase, $Cd$) 
is shorter in these SNe. Hence, we may expect a relation between $Cd$ and progenitor radius.
Moreover, \citet{Morozova16} found that the early properties of the light 
curve are sensitive to the progenitor radius, which implies that the rise time has a relation with the radius at the time
of the explosion. \citet{Gonzalez15} using a large sample of observed SNe~II, concluded that SNe~II progenitor
radii are relatively small.
We note however the recent results of \citet{Yaron17}, \citet{Morozova17}, \citet{Moriya17} and \citet{Dessart17}.
These investigations have provided evidence for and shown the effect of previously unaccounted for material close to the progenitor star. 
The interaction of the SN ejecta with such material may thus complicate the relation between early-time observations and 
progenitor radius.\\
\indent In summary we expect that \textit{the hydrogen envelope mass} is directly 
related with $Pd$, s$_{3}$; \textit{the explosion energy}  with the expansion velocities ($vel$), 
and the luminosities (M$_{max}$, M$_{end}$); and \textit{the radius of the progenitor}
would have some influence in $Cd$.

\section{Results}
\label{results}

In this section we investigate the spectral and photometric diversity of SNe~II through correlations.
Here we present the statistics of these correlations and their
respective figures. As stated above, the spectral measurements were performed in the phases where 
the data were available, however to characterize this diversity, the analysis is done at 30, 50 and 80 days 
with respect to the explosion epoch. 
In Table~\ref{average} we can see the average of the correlations for each parameter at 30, 50 and 80 days.
The mean of these correlations shows a value of 0.323, 0.364 and 0.356 for each epoch, thus the following analysis
is performed at 50 days, where more spectral measurements are available and the mean is higher.     
In Tables~\ref{velocidades}, and~\ref{pews} the measured spectral parameters  at 50 days are listed,
while in Table~\ref{photo} we present the photometric parameters.

\subsection{Spectral correlations in the photospheric phase}
\label{specor}

\begin{figure}
\centering
\includegraphics[width=8.2cm]{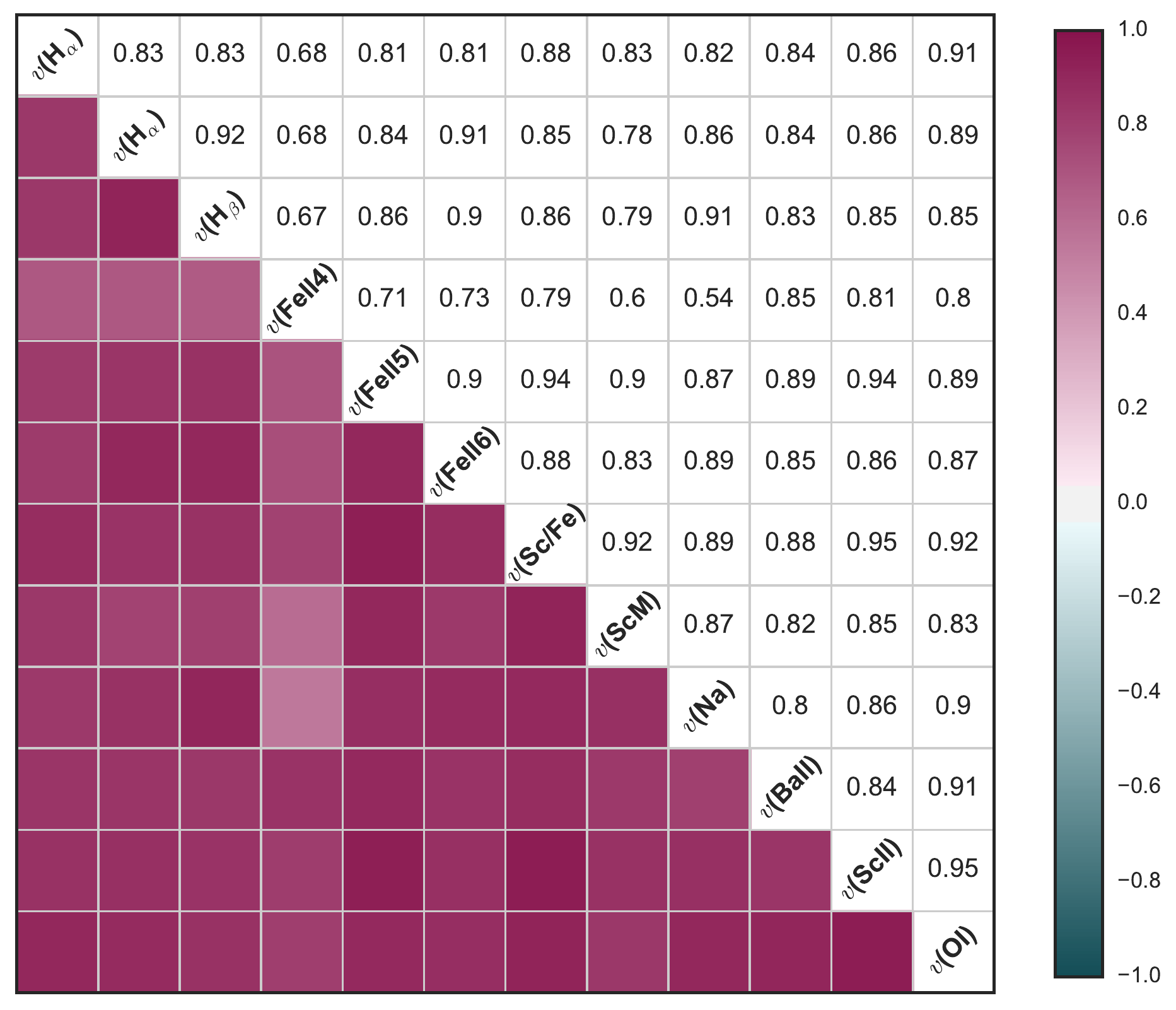}
\caption{Correlation matrix of the individual velocity measurements at 50 days.
Colors indicate the Pearson correlation coefficient $\rho$. The diagonal middle line shows the name of the 
parameter: H$_{\alpha}$ from FWHM and from the minimum absorption flux, H$_{\beta}$, \ion{Fe}{2} $\lambda4924$, \ion{Fe}{2} $\lambda5018$,
\ion{Fe}{2} $\lambda5169$, \ion{Sc}{2}/\ion{Fe}{2} $\lambda5531$, \ion{Sc}{2} M $\lambda5663$, \ion{Na}{1} D, \ion{Ba}{2} $\lambda6142$, \ion{Sc}{2} $\lambda6247$, and
\ion{O}{1} $\lambda7774$ velocities.}
\label{vel50}
\end{figure}

We analyze the spectral properties of SNe~II, focusing on correlations between pEWs, expansion velocities,
velocity decline rate, and velocity differences. Figure~\ref{vel50} shows the correlation matrix
of the velocity measurements at 50 days obtained by estimating the Pearson correlation coefficient.
Correlation coefficients are displayed in color: darkest colors (green and purple) represent the highest
correlation found with the Pearson correlation test (-1 and 1, respectively), while white colors (0)
mean no correlation. These colors are presented in the lower triangle, while 
the upper triangle shows the Pearson correlation value  ($\rho$). 
It is generally considered that correlation coefficients between 0 and 0.19 represent 
close to zero correlation, 0.2-0.39 weak, 0.4-0.59 moderate, 0.6-0.89 strong, and 0.8-1.0 very strong 
\citep{Evans96}, while also noting the statistical significance of these correlation coefficients
in many cases.
We will use these descriptions for the following discussion.
As shown in Figure~\ref{vel50},  all velocities strongly correlate positively with each other, 
as we would expect for an homologous expansion ($v \propto r$). 
Taking an average, $v$(\ion{Sc}{2}/\ion{Fe}{2}) $\lambda5531$, $v$(\ion{O}{1}) $\lambda7774$ and $v$(\ion{Sc}{2}) $\lambda6247$ 
show the highest correlations with the other parameters, with values of 0.887, 0.883 and 0.875, respectively,
while \ion{Fe}{2} $\lambda4924$ shows the lowest (0.714).
The \ion{Sc}{2} $\lambda6247$ line velocities correlate strongly with \ion{Fe}{2} $\lambda5018$ and \ion{Sc}{2}/\ion{Fe}{2} $\lambda5531$, 
with a value of $\rho=0.94$ and $\rho=0.95$.
It is important to note that while the velocities all correlate, they are offset. In general, 
the differences in the velocities are related to the optical depth for each line and the proximity of the 
line forming region to the photosphere. As H$_{\alpha}$ displays the highest velocities, it is mostly formed in the
outer shell of the ejecta and its optical depth is much larger than the \ion{Fe}{2} lines, which
are forming near to the photosphere.

\begin{figure}
\centering
\includegraphics[width=8.2cm]{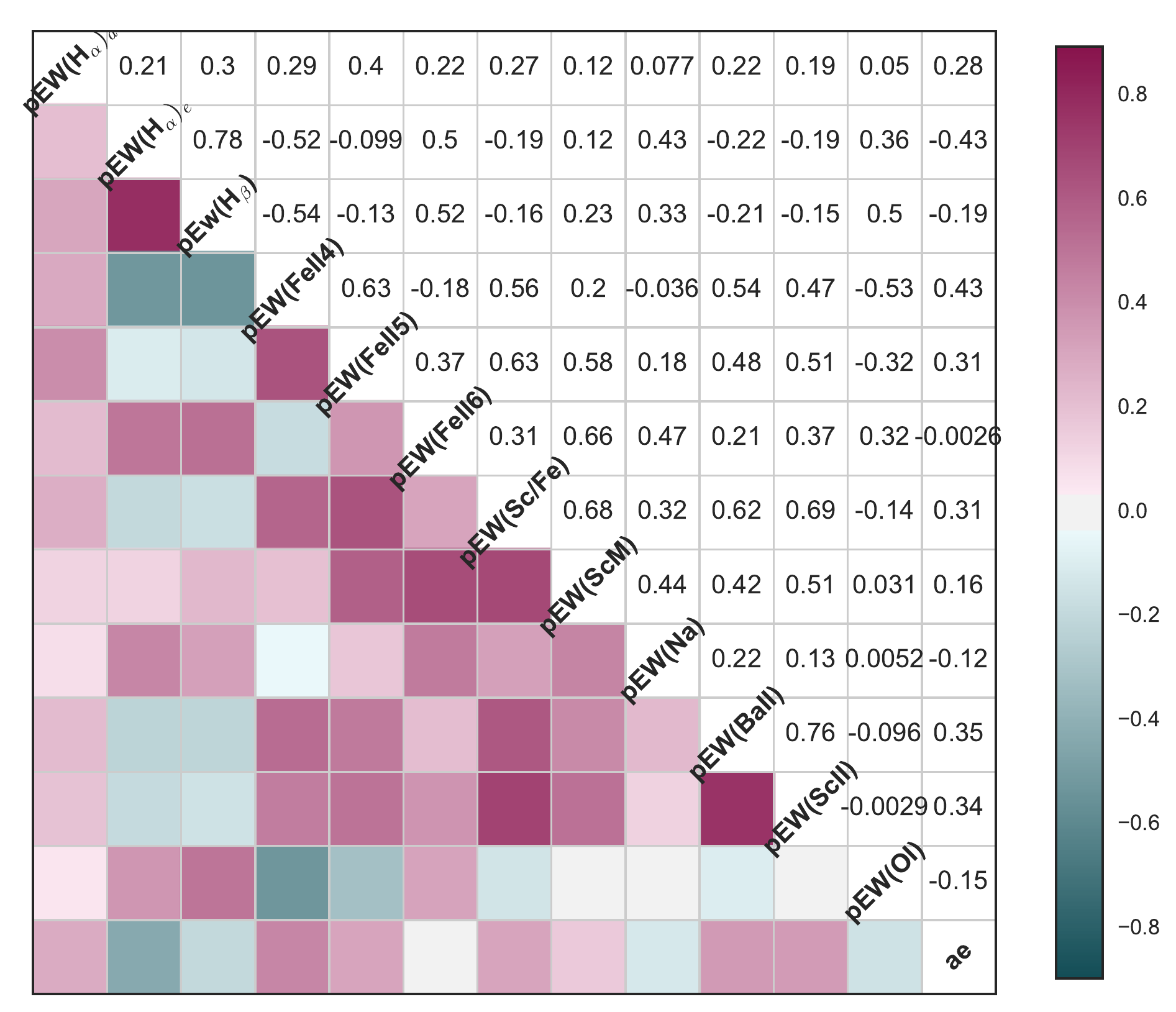}
\caption{Correlation matrix of the individual pEW measurements at 50 days.
Colors indicate the Pearson correlation coefficient $\rho$. The diagonal middle line shows the name of the 
parameter: pEW(H$_{\alpha}$) of absorption component, pEW(H$_{\alpha}$) of emission component, pEW(H$_{\beta}$), 
pEW(\ion{Fe}{2} $\lambda4924$), pEW(\ion{Fe}{2} $\lambda5018$), pEW(\ion{Fe}{2} $\lambda5169$), pEW(\ion{Sc}{2}/\ion{Fe}{2} $\lambda5531$), 
pEW(\ion{Sc}{2} M $\lambda5663$), pEW(\ion{Na}{1} D), pEW(\ion{Ba}{2} $\lambda6142$), pEW(\ion{Sc}{2} $\lambda6247$), pEW(\ion{O}{1} $\lambda7774$) and $a/e$.}
\label{pew50}
\end{figure}

\indent Figure~\ref{pew50} shows the correlation matrix of the pEWs measurements at 50 days. 
Searching for correlations of pEWs with each other, we find that \ion{Sc}{2}/\ion{Fe}{2} $\lambda5531$ 
seems to be the dominant parameter to correlate with all the other pEWs (on average 0.404), while
the pEW of H$_{\alpha}$ absorption component shows very weak correlations with other pEWs.
The strongest correlations are displayed by the iron-group lines with each other. 
We can see moderate correlations between the pEW of \ion{O}{1} $\lambda7774$ and H$_{\beta}$. 
In the case of $a/e$ we find a moderate correlation only with \ion{Fe}{2} $\lambda4924$ ($\rho=0.43$) and 
anticorrelation with pEW of H$_{\alpha}$ emission ($\rho=-0.43$).
While H$_{\beta}$ shows a weak correlation with the H$_{\alpha}$ absorption component ($\rho=0.3$),
the correlation with the H$_{\alpha}$ emission component is strong, with a  $\rho=0.78$. 
The lack of correlation between H$_{\alpha}$ and H$_{\beta}$ absorption features could be due to a)blending 
effects of \ion{Fe}{2}, \ion{Sc}{2} and \ion{Ba}{2} lines with H$_{\beta}$, and/or b) the effects of 
Cachito (Paper I) on the profile of H$_{\alpha}$.

\begin{figure*}
\centering
\includegraphics[width=\textwidth]{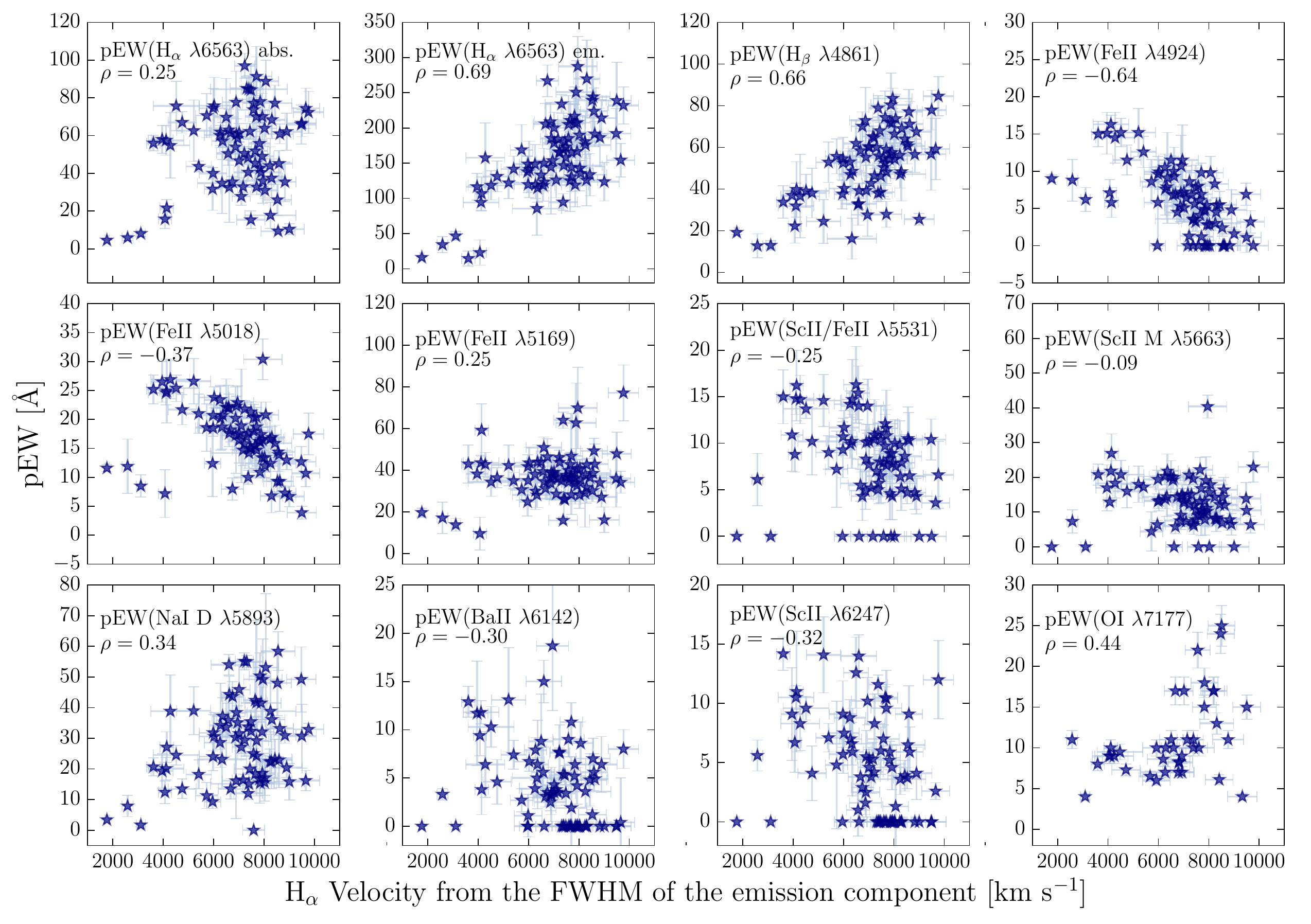}
\caption{Relations between H$_{\alpha}$ velocities and the
pEWs of H$_{\alpha}$ of absorption and emission
component, H$_{\beta}$, \ion{Fe}{2} $\lambda4924$, \ion{Fe}{2} $\lambda5018$, \ion{Fe}{2} $\lambda5169$, \ion{Sc}{2}/\ion{Fe}{2} $\lambda5531$, 
\ion{Sc}{2} multiplet, \ion{Na}{1} D, \ion{Ba}{2} $\lambda6142$, \ion{Sc}{2} $\lambda6247$, and \ion{O}{1} $\lambda7774$.
On the top left of each panel the spectral feature name is displayed, together with the Pearson correlation value.}
\label{comp1}
\end{figure*}

\begin{figure*}
\centering
\includegraphics[width=\textwidth]{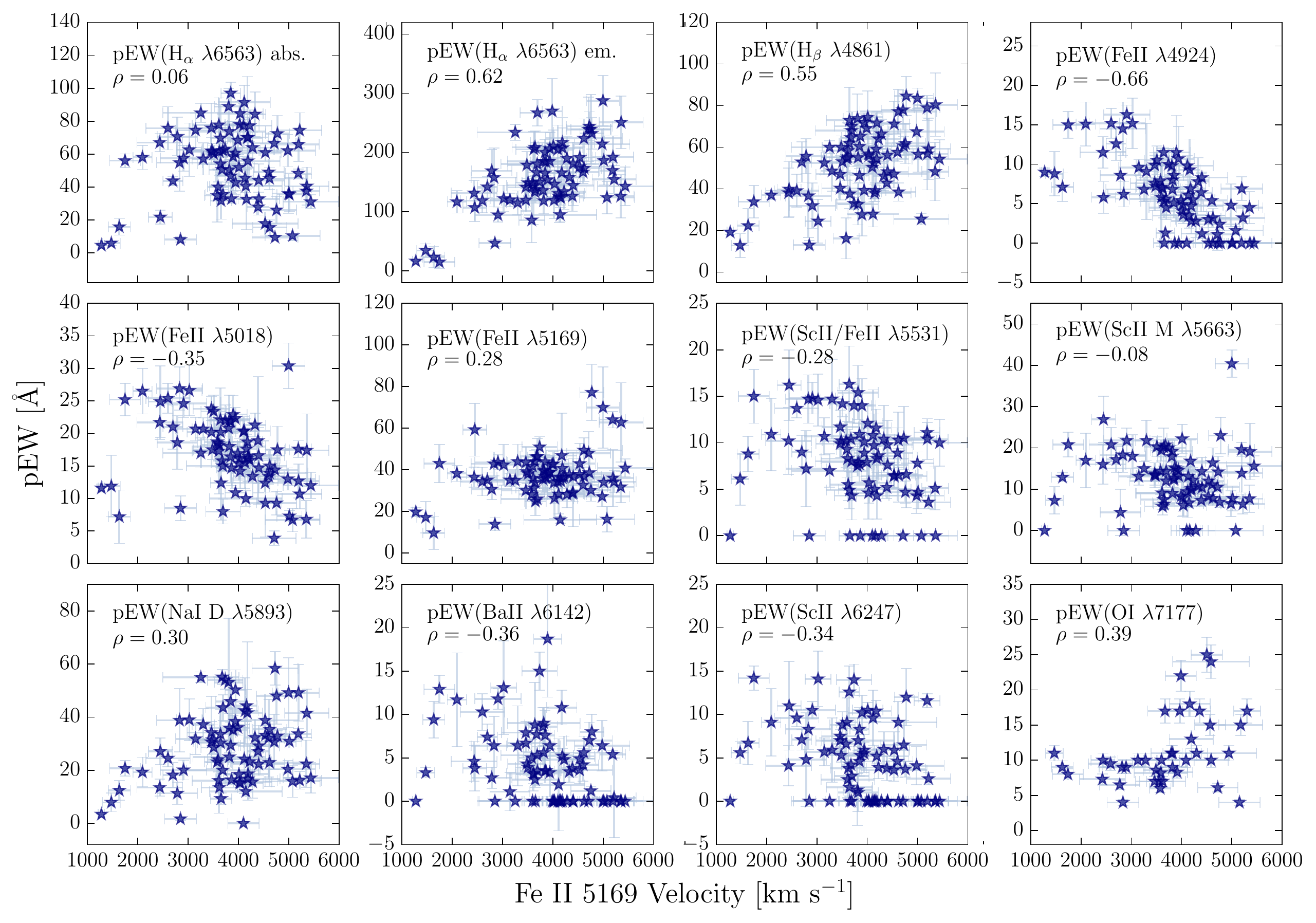}
\caption{Same as Figure~\ref{comp1} but for \ion{Fe}{2} 5169 velocities.}
\label{comp2}
\end{figure*}

\begin{figure*}
\centering
\includegraphics[width=\textwidth]{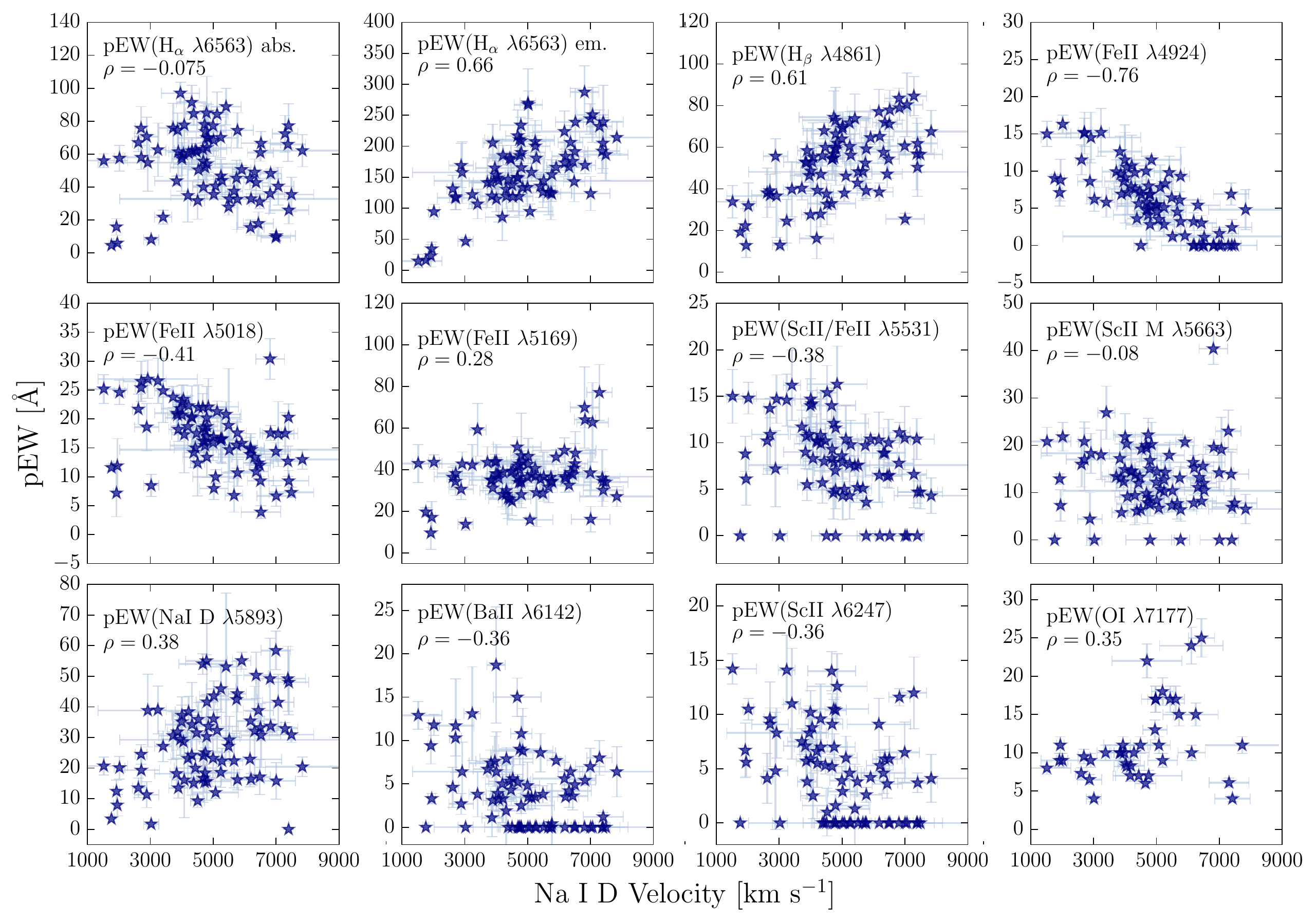}
\caption{Same as Figure~\ref{comp1} but for \ion{Na}{1} D velocities.}
\label{comp3}
\end{figure*}

\indent Figures~\ref{comp1}, \ref{comp2}, and \ref{comp3}, show the relations between the
H$_{\alpha}$,  \ion{Fe}{2} $\lambda5169$, and  \ion{Na}{1} D velocities and the pEWs 
for the 11 features explained above at 50 days. Checking these correlations we see that velocities
correlate positively with Balmer and \ion{Na}{1} D lines, but negatively with \ion{Fe}{2} lines.
For H$_{\alpha}$ we present the pEW of the absorption and emission component in the first two panels, respectively.
In the three figures are shown five objects with the lowest velocities and smallest 
pEW values. Three of these SNe show signs of interaction (narrow emision lines) at early 
times (SN~2008bm, 2009au and 2009bu, these SNe also display abnormally low velocities for their brightness). 
The other two SNe are SN~2008br and SN~2002gd. 
In those panels plotting pEWs of \ion{Fe}{2} $\lambda4924$, \ion{Sc}{2}/\ion{Fe}{2} $\lambda5531$, 
\ion{Sc}{2} $\lambda5663$, \ion{Ba}{2} $\lambda6142$, and \ion{Sc}{2} $\lambda6247$, one can see that there are many SNe with $pEW=0$. 
In these spectra we do not detect these lines.\\
\indent In Figure~\ref{comp1} we can see that the H$_{\alpha}$ velocities do not show correlation with 
pEW(H$_{\alpha}$) of the absorption component, pEW(\ion{Fe}{2} $\lambda5169$), pEW(\ion{Sc}{2}/\ion{Fe}{2} $\lambda5531$), pEW(\ion{Sc}{2}
multiplet), pEW(\ion{Na}{1} D), pEW(\ion{Ba}{2} $\lambda6142$), and pEW(\ion{Sc}{2} $\lambda6247$).
The strongest correlations are shown 
with pEW(H$_{\alpha}$) of the emission component, H$_{\beta}$, and  anticorrelations with \ion{Fe}{2} $\lambda4924$,
and \ion{Fe}{2} $\lambda5018$. Figures~\ref{comp2} and \ref{comp3} show that \ion{Fe}{2} $\lambda5169$ and \ion{Na}{1} D velocities
present more scatter in their relations than those shown by H$_{\alpha}$ velocities.

The expansion velocities with $\Delta v$(H$_{\beta})$ show anticorrelations, which are stronger 
at late epochs (between 50 and 80 days) than at early phases (15 to 30 days, 15 to 50 days,
and 30 to 50 days). 
Meanwhile, $\Delta vel$(H$_{\alpha}-$\ion{Fe}{2} $\lambda5018$) and $\Delta vel$(\ion{Na}{1} D$-$\ion{Fe}{2} $\lambda5018$)
show correlations with the expansion velocities at 50 days (see Figure~\ref{pearson50}).

\subsection{Spectroscopic and photometric properties}
\label{correl}

We now present a comparison of spectroscopic and photometric properties of SNe~II.
While we have defined and measured 31 spectroscopic and 13 photometric parameters, 
here we choose a smaller number of parameters to focus on and search for correlations between them. Thus, we employ 
14 spectral and 11 photometric parameters: $v$(H$_{\alpha}$) obtained from the FWHM of the emission component,
$v$(H$_{\beta}$), $v$(\ion{Fe}{2} 5018), $v$(\ion{Fe}{2} 5169), $v$(\ion{Na}{1} D), pEW(H$_{\alpha(abs)}$),
pEW(H$_{\alpha(emis)})$, pEW(H$_{\beta)}$), pEW(\ion{Fe}{2} 5018), pEW(\ion{Fe}{2} 5169), pEW(\ion{Na}{1} D),
$a/e$, $\Delta v(H_{\beta})$ in a range of $50\leq t \leq80$d, $\Delta vel$(H$_{\alpha}-$\ion{Fe}{2} 5018), 
$\Delta vel$(\ion{Na}{1} D$-$\ion{Fe}{2} 5018), $OPTd$, $Pd$, $Cd$, M$_{max}$, M$_{end}$, M$_{tail}$, s$_1$, s$_2$, 
s$_3$, $\Delta {(B-V)}$ in a range of $10\leq t \leq30$ d, and the $^{56}$Ni mass.

\begin{figure*}
\centering
\includegraphics[width=18cm]{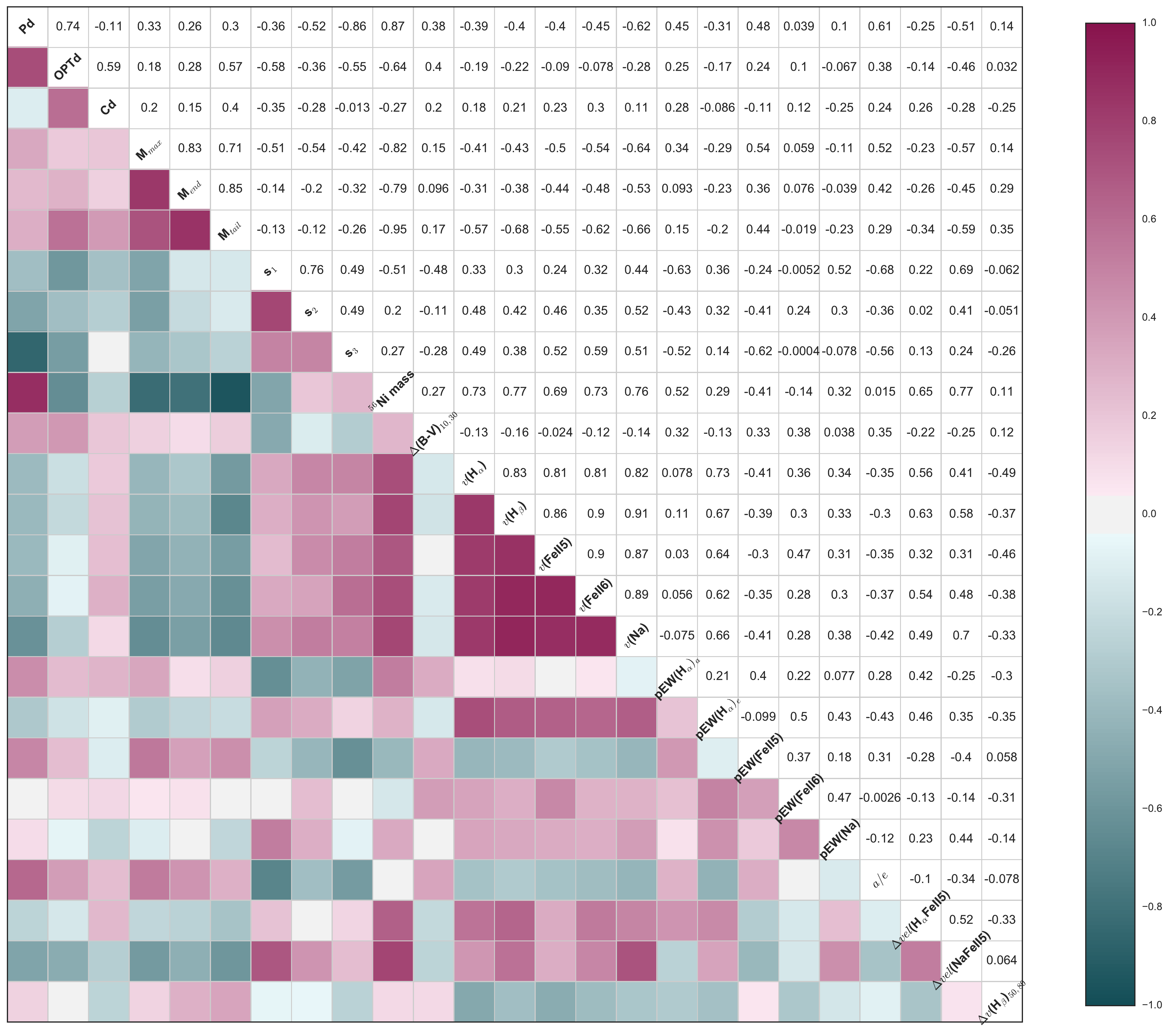}
\caption{Correlation matrix of the individual spectral and photometric  parameters at 50 days.
Colors indicate the Pearson correlation coefficient $\rho$. In the diagonal line is shown
$Pd$: plateau duration; $OPTd$: optically thick duration; $Cd$: cooling duration; M$_{max}$: magnitude at maximum;
M$_{end}$: magnitude at the end of the plateau; M$_{tail}$: magnitude in the radioactive tail phase; s$_1$: initial decline; 
s$_2$: plateau decline; s$_3$: radioactive tail decline; $^{56}$Ni mass: nickel mass; $\Delta {(B-V)}_{10,30}$: color gradiente
between 10 and 30 days from explosion; $v$(H$_{\alpha}$): H$_{\alpha}$ velocity obtained from the FWHM of the emission component;
$v$(H$_{\beta}$): H$_{\beta}$ velocity; $v$(FeII5): \ion{Fe}{2} 5018 velocity; $v$(FeII6): \ion{Fe}{2} 5169 velocity; $v$(Na):
\ion{Na}{1} D velocity, pEW(H$_{\alpha})_a$: pEW of H$_{\alpha}$ absorption component; pEW(H$_{\alpha})_e$: pEW of the 
H$_{\alpha}$ emission component, pEW(H$_{\beta)}$: pEW of H$_{\beta}$, pEW(FeII5): pEW of \ion{Fe}{2} 5018, pEW(FeII6): pEW of 
\ion{Fe}{2} 5169; pEW(Na): pEW of\ion{Na}{1} D, $a/e$: ratio of absortion to emission component of H$_{\alpha}$ P-Cygni profile;
$\Delta vel$(H$_{\alpha}$FeII5): $\Delta vel$(H$_{\alpha}-$\ion{Fe}{2} 5018), $\Delta vel$(NaFeII5): 
$\Delta vel$(\ion{Na}{1} D$-$\ion{Fe}{2} 5018); and
$\Delta v(H_{\beta})_{50,80}$: $\Delta v(H_{\beta})$ in a range of $[+50,+80]$ days. 
}
\label{pearson50}
\end{figure*}

\begin{figure}
\centering
\includegraphics[width=\columnwidth]{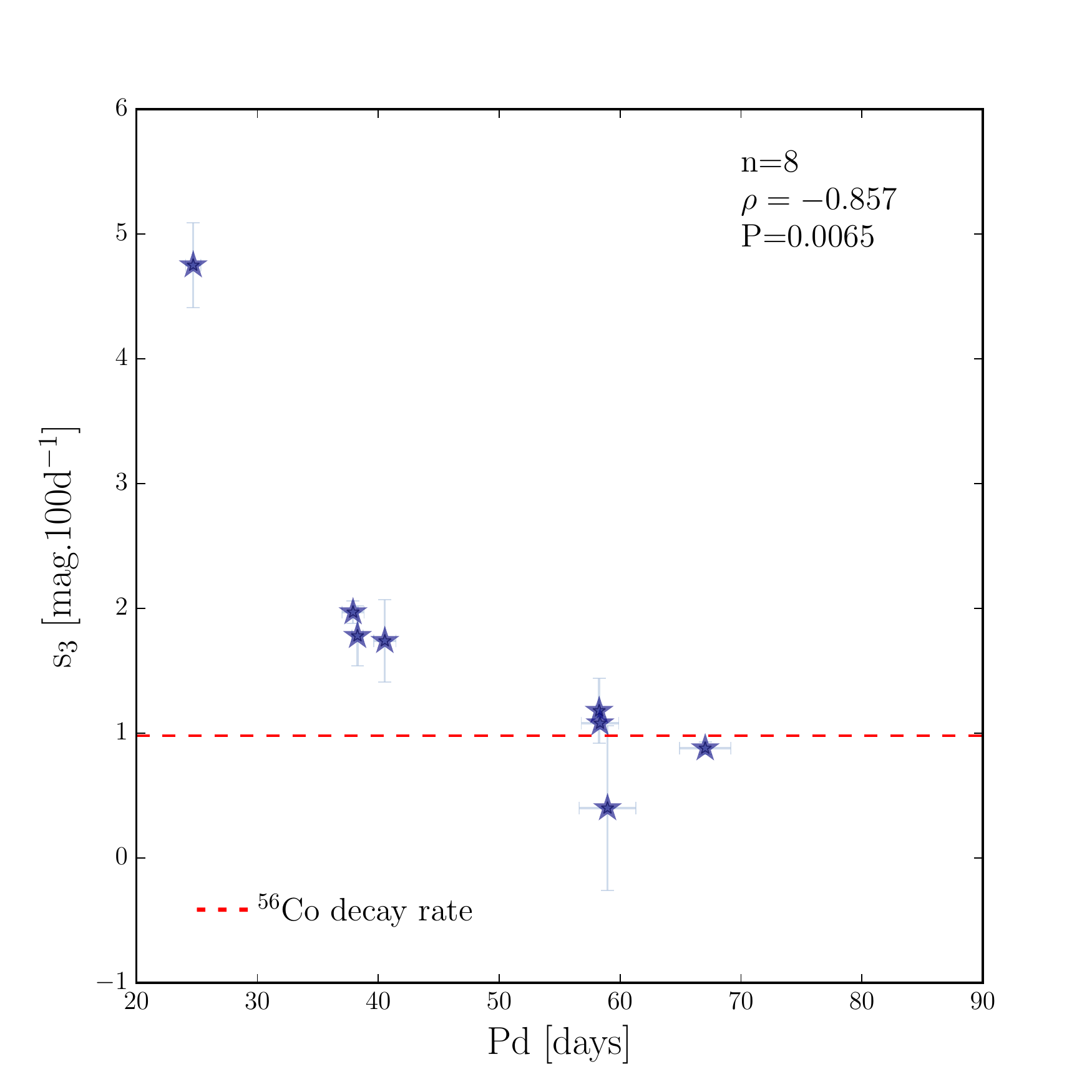}
\caption{Correlation between $Pd$ vs. s$_3$. On the top of the figure: n = number of events, $\rho$ = Pearson’s
correlation coefficient, and P = probability of detecting a correlation by chance.
The dashed horizontal line shows the expected decline rate on the radioactive tail, assuming full trapping of
gamma-rays from $^{56}$Co to $^{56}$Fe decay.}
\label{pds3}
\end{figure}

Figure~\ref{pearson50} shows the correlation matrix of the spectroscopic parameters (obtained at 50 days from explosion)
and photometric properties. Although photometric correlations have been shown in 
previous works \citep[e.g. A14,][]{Valenti16}, the incorporation of numerous spectral parameters
can aid in furthering our understanding of the link between observed parameters and underlying SN~II physics.
As in the previous matrix of correlation, darkest colors indicate higher 
correlation and white colors, no correlation. \\
\indent Focusing on the photometric correlations, one can see that many of these are stronger
than in A14. As discussed previously, this is because some parameters have been remeasured with 
new techniques (Galbany et al. in prep). Interestingly, the number of SNe~II with measured values of both
$Pd$ and s$_3$ show an increase from 4 in A14
to 8 in this work.  As explained above, both parameters can give us an idea of the of hydrogen
envelope mass at the moment of explosion, thus some relation is expected. 
Figure~\ref{pds3} shows an evident trend between both parameters, with a correlation coefficient 
of $\rho=-0.857$ (although we note the low number of SNe). SNe~II with smaller $Pd$ have higher s$_3$ decline rates, providing further evidence of a dominant
role in defining light-curve morphology of the hydrogen envelope mass, while also providing further support for
the use of $Pd$ and s$_3$ as envelope mass indicators (given their relatively strong correlation).

\begin{figure}
\centering
\includegraphics[width=\columnwidth]{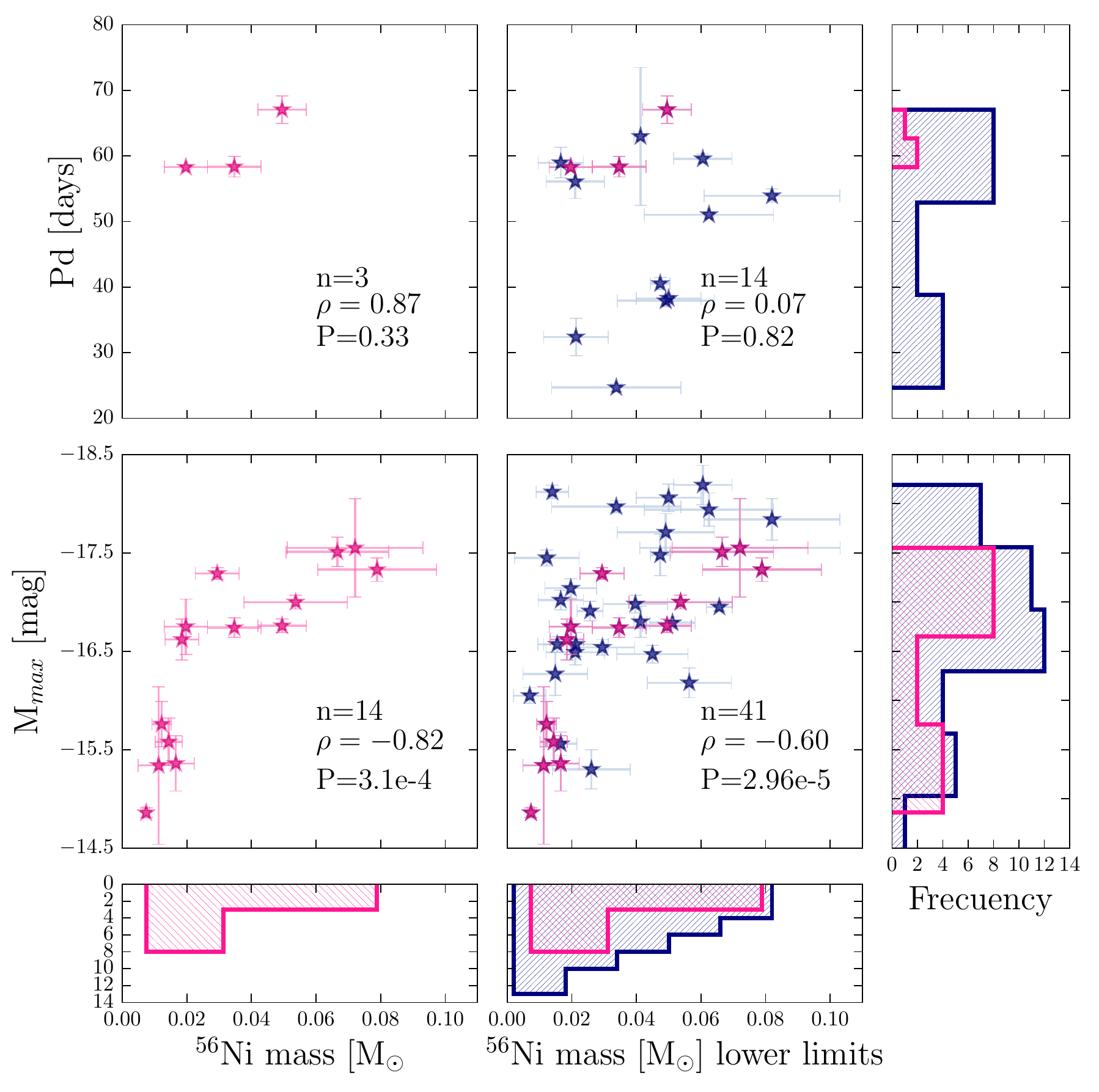}
\caption{\textit{Top:} Correlations between $Pd$ and the $^{56}$Ni mass with the accurate values (left)
and including the lower limits (right). 
\textit{Bottom:} Correlations between M$_{max}$ and the $^{56}$Ni mass with the accurate values (left)
and including the lower limits (right). The accurate values for $^{56}$Ni mass are display in red. 
On the top of each figure: n = number of events, $\rho$ = Pearson’s
correlation coefficient, and P = probability of detecting a correlation by chance. Histograms
along the x and y-axes show the distributions of the various parameters plotted on each axis.
Each histogram displays the $\rho_{val}$ found using the Shapiro-Wilk normalization. When the $\rho_{val}$ $>0.05$,
the dataset comes from a population which has a normal distribution.}
\label{niq}
\end{figure}

\indent From Figure~\ref{pearson50} we also can see that $Pd$ has a moderate correlation with 
velocities.
Although we find a strong correlation between $Pd$ and $^{56}$Ni mass, in agreement to
the theoretical predictions \citep[e.g.][]{Kasen09}, we are not in a position to support this result
because the correlation is produced only with three points.
However, when we include the lower limits for the $^{56}$Ni mass,
the correlation disappears (see top panel in Figure~\ref{niq}). In general, the correlations between the 
$^{56}$Ni mass and all other parameters decrease when we use the lower limits. In the bottom panel of the same
plot (Figure~\ref{niq}) it is possible to see how the scatter increases using the these values. The correlation 
goes from $\rho=-0.82$ to $\rho=-0.60$. The fact that correlations become weaker when using lower 
$^{56}$Ni mass limits suggests that one should be careful analysing such masses when insufficient data are available 
for their estimation.

\begin{figure*}
\includegraphics[width=\textwidth]{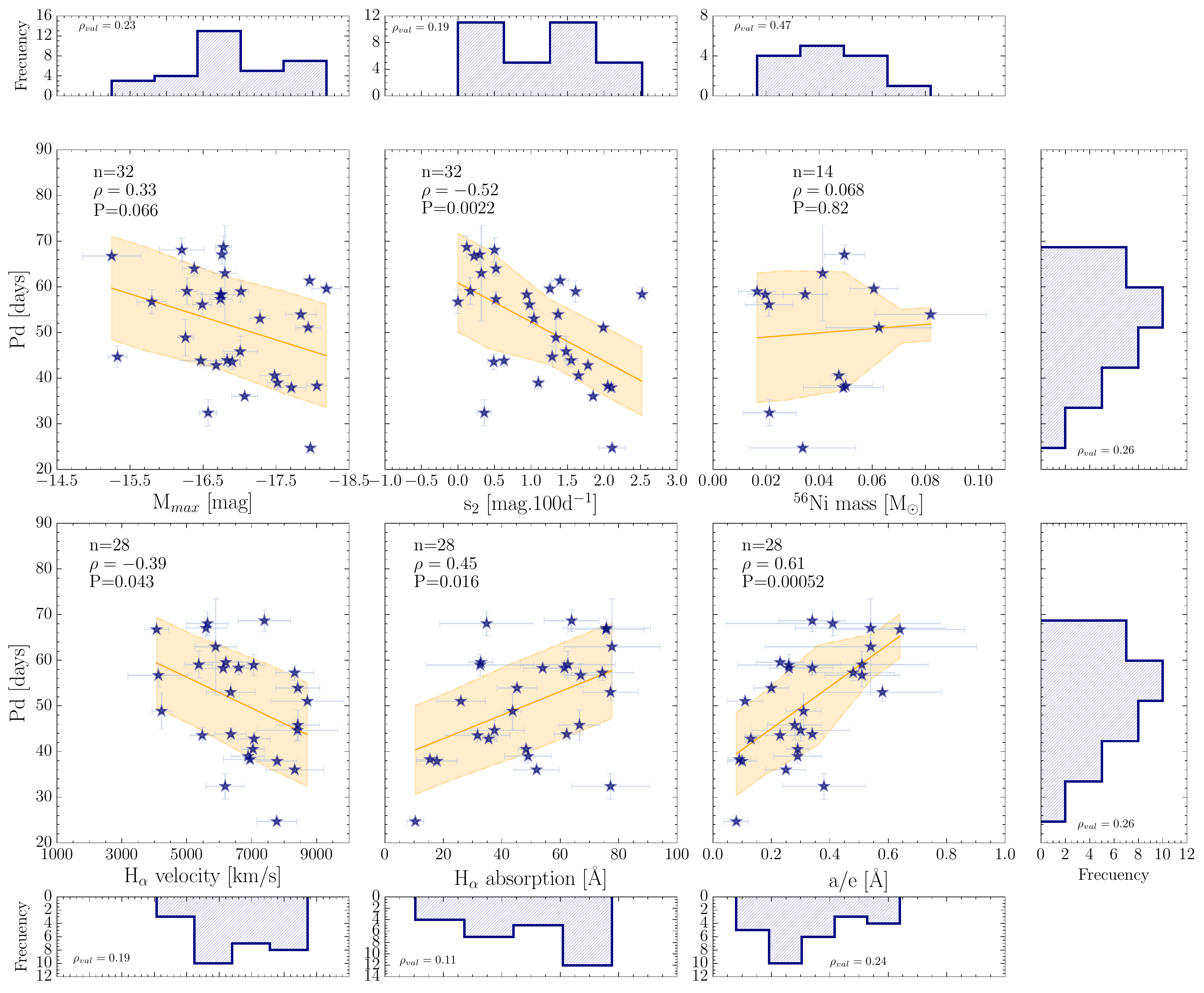}
\caption{Correlations between $Pd$ and six different parameters:
M$_{max}$, s$_2$, $^{56}$Ni mass, H$_{\alpha}$ velocity, pEW of H$_{\alpha}$ absorption component, $a/e$.
On the top of the figure: n = number of events, $\rho$ = Pearson’s
correlation coefficient, and P = probability of detecting a correlation by chance. In addition, each plot
shows the corresponding best fit (\texttt{linmix\_err}; \citealt{Kelly07}) as solid orange line, while the orange shaded
area indicates the variance with
respect to the fit line. Histograms along the x and y-axes show the distributions of the various parameters plotted on each axis.
Each histogram displays the $\rho_{val}$ found using the Shapiro-Wilk normalization. When the $\rho_{val}$ $>0.05$,
the dataset comes from a population which has a normal distribution.}
\label{pd}
\end{figure*}

\indent Continuing the analysis of $Pd$, we can see that it has a moderate correlation with 
pEW(H$_{\alpha}$) of the absorption component and strong correlation with $a/e$.
The correlation coefficients are $\rho=0.45$ and $\rho=0.61$, 
respectively. In Figure~\ref{pd} we present these correlations together with the best fit line obtained 
using the \texttt{linmix\_err}\footnote{A Bayesian approach to linear regression with errors in both X and Y.} package
\citep{Kelly07} and the variance with respect to the fit line. The trend shows that SNe with shorter $Pd$ values
are brighter, have faster declining light curves, lower pEW(H$_{\alpha}$) of the absorption component and $a/e$ values,
and higher velocities, however the scatter is large. In many cases this scatter is significantly larger than 
the that which could be ascribed to the errors on individual data points. This suggests that this scatter is due to  
differing underlying physics driving diversity in different parameters plotted on each axis. For example, while we 
argue here that $Pd$ is a good indicator of the hydrogen envelope mass, theory also predicts this parameter to be 
influenced by the $^{56}$Ni mass \citep{Kasen09}. Meanwhile, SN luminosities and velocities will be affected by both 
explosion energy and the ejecta/envelope mass. Interaction of the SN ejecta with CSM material at early times 
\citep[e.g.][]{Morozova17,Moriya17,Dessart17} may also play a role in producing dispersion in our presented trends.\\
\indent The fact that we see a significant anti-correlation between $Pd$ and s$_2$ is in line with historical
understanding of the nature of fast declining SNe~II. If $Pd$ is an indicator of the extent of the
hydrogen envelope, then it follows that faster declining SNe~II have a smaller hydrogen envelope at the epoch of 
explosion, consistent with previous theoretical predictions \citep[e.g][]{Popov93, Litvinova83, Bartunov92, Moriya15}.

\begin{figure*}
\centering
\includegraphics[width=\textwidth]{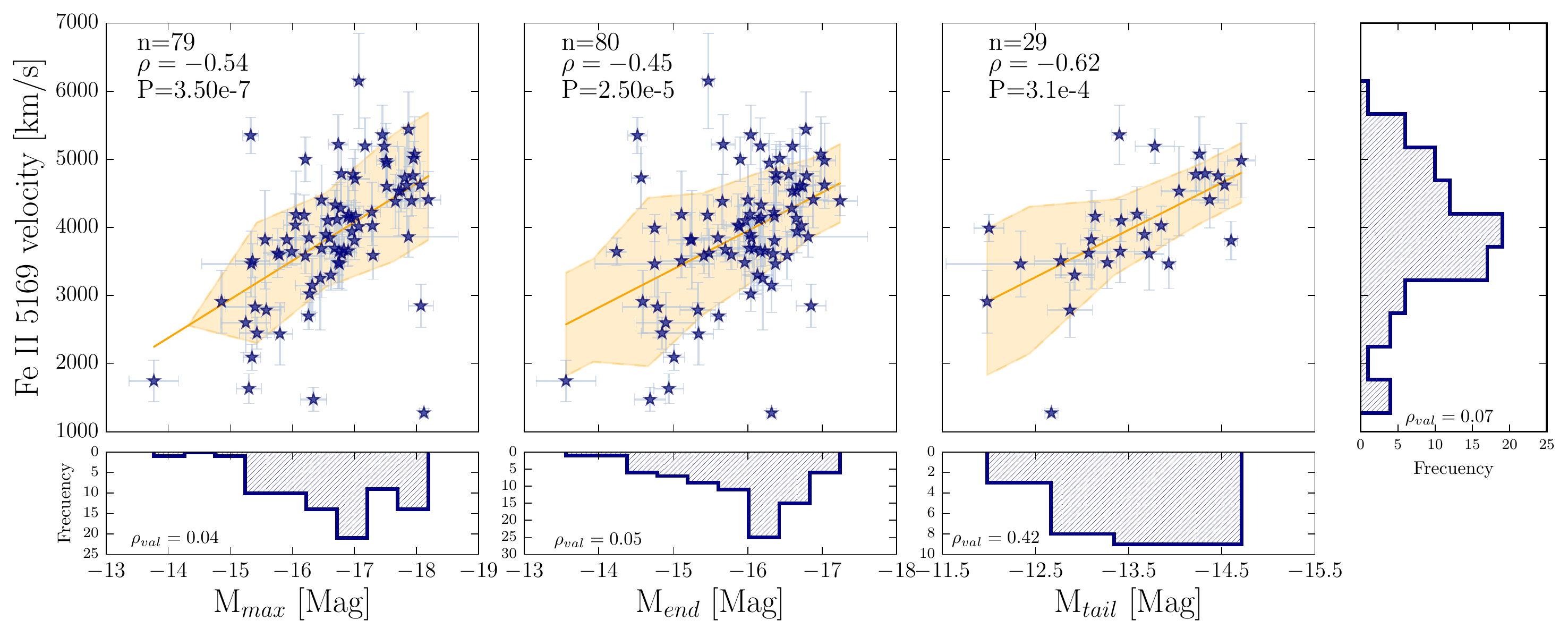}
\caption{Correlations between (Fe II $\lambda$5169) velocity and the magnitudes: M$_{max}$, M$_{end}$ and M$_{tail}$.
In the top left of each plot the following values are given: n = number of events, $\rho$ = Pearson’s
correlation coefficient, and P = probability of detecting a correlation by chance. In addition, each plot
shows the corresponding best fit (\texttt{linmix\_err}; \citealt{Kelly07}) as solid orange line, while the orange shaded
area indicates the variance with
respect to the fit line. Histograms along the x and y-axes show the distributions of the various parameters plotted on each axis.
Each histogram displays the $\rho_{val}$ found using the Shapiro-Wilk normalization. When the $\rho_{val}$ $>0.05$,
the dataset comes from a population which has a normal distribution.}
\label{femag}
\end{figure*}

\indent In Figure~\ref{femag} we test the correlation found by \citet{Hamuy02L} between the magnitude and the
photospheric expansion velocity. Unlike \citet{Hamuy02L}, who only used SNe~IIP and the $M_{\rm V}$ in the middle
of the plateau, we use all our SN~II sample (no distinction between SNe~IIP and SNe~IIL)
and the magnitude at different phases: at maximum (M$_{max}$), at the end of the plateau (M$_{end}$) and at the 
radioactive tail phase (M$_{tail}$). We can see that brighter events (in all phases) display higher expansion velocities,
confirming the result of \citet{Hamuy02L}. The correlations between \ion{Fe}{2} $\lambda5169$ velocity (a proxy of the photospheric 
velocity) at 50 days and luminosity during the optically thick phase are moderate ($\rho=-0.54$ with M$_{max}$ and $\rho=-0.45$ with 
M$_{end}$), and strong ($\rho=-0.62$) in the radioactive tail phase. However, we again note the outliers in these figures, 
where the correlation appears much stronger when removing these events (the outliers are mainly the same SNe 
discussed previously that show abnormal spectral properties). Interestingly, correlations are higher between spectral velocities
and M$_{max}$ than with M$_{end}$ (the Standardized Candle Method, SCM, is generally applied using a magnitude during the plateau,
more similar to M$_{end}$). Analysing the variance along the best fit line, we find that the dispersion in velocity is
larger in brighter SNe. Although the magnitudes and the expansion velocities are both directly related with the explosion energy, 
this scatter could suggest an extra influence by an external parameter. In the three main outliers in this plot we observe signs
of weak interaction at early times (see spectra presented in Paper I). In these three obvious cases, but also
in other more `normal' SNe~II, interaction could play a role in influencing both the magnitudes and velocities observed. 
CSM interaction is likely to produce more dispersion within brighter SNe~II as it will generally increase the early-time luminosity
while possible decreasing velocities, hence pushing SNe~II away from the classic magnitude-velocity relation.
In addition, the unaccounted for effects of host galaxy reddenning will produce additional dispersion.

\begin{figure*}
\centering
\includegraphics[width=\textwidth]{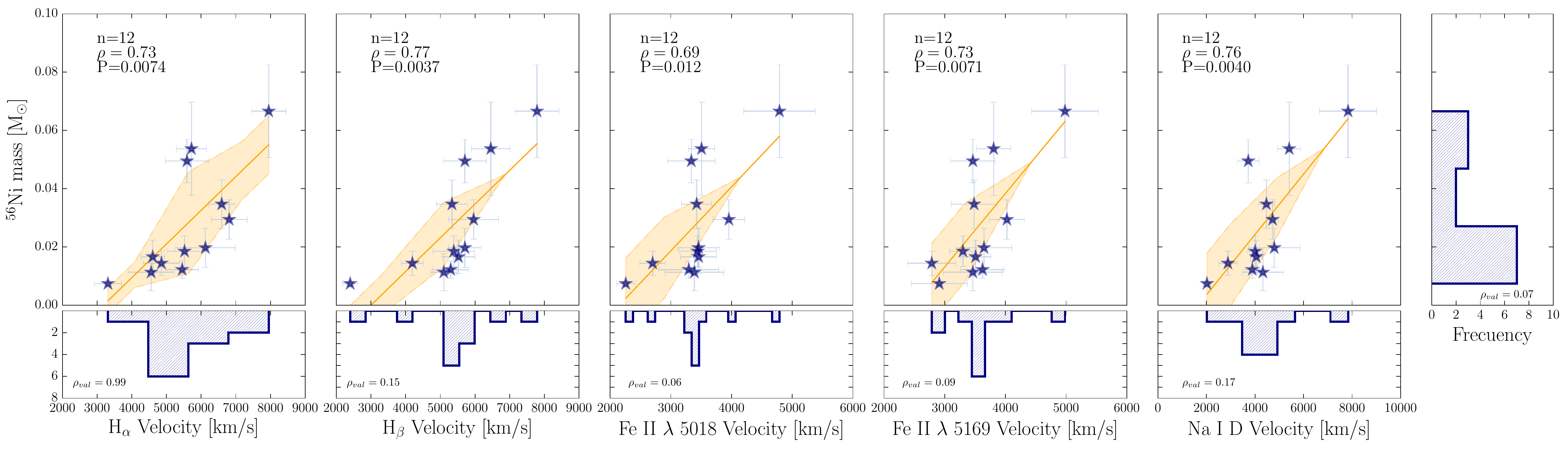}
\caption{Correlations between $^{56}$Ni and the expansion velocities.
On the top of the figure: n = number of events, $\rho$ = Pearson’s
correlation coefficient, and P = probability of detecting a correlation by chance. In addition, each plot
shows the corresponding best fit (\texttt{linmix\_err}; \citealt{Kelly07}) as solid orange line, while the orange shaded
area indicates the variance with
respect to the fit line. Histograms along the x and y-axes show the distributions of the various parameters plotted on each axis.
Each histogram displays the $\rho_{val}$ found using the Shapiro-Wilk normalization. When the $\rho_{val}$ $>0.05$,
the dataset comes from a population which has a normal distribution.}
\label{velni}
\end{figure*}

\indent The expansion velocities show a strong correlation with $^{56}$Ni mass 
(see Figure~\ref{velni}). This suggests that more energetic explosions produce more 
$^{56}$Ni. Additionally, the luminosities have a very strong correlation with the $^{56}$Ni mass,
which supports the results obtained by \citet{Hamuy03,Pejcha15,Pejcha15a} and more recently by \citet{Muller17}.
It is possible to see that these three parameters (luminosities, velocities and $^{56}$Ni mass) 
are related and they can be explained through a correlation of both parameters with explosion energy: 
more energetic explosions produce brighter SNe with faster velocities 
(as shown in the models of \citealt{Dessart10b}). 
For those correlations that we do not plot, the reader can see the strength of correlation in Figure~\ref{pearson50}.

\begin{figure*}
\centering
\includegraphics[width=14cm]{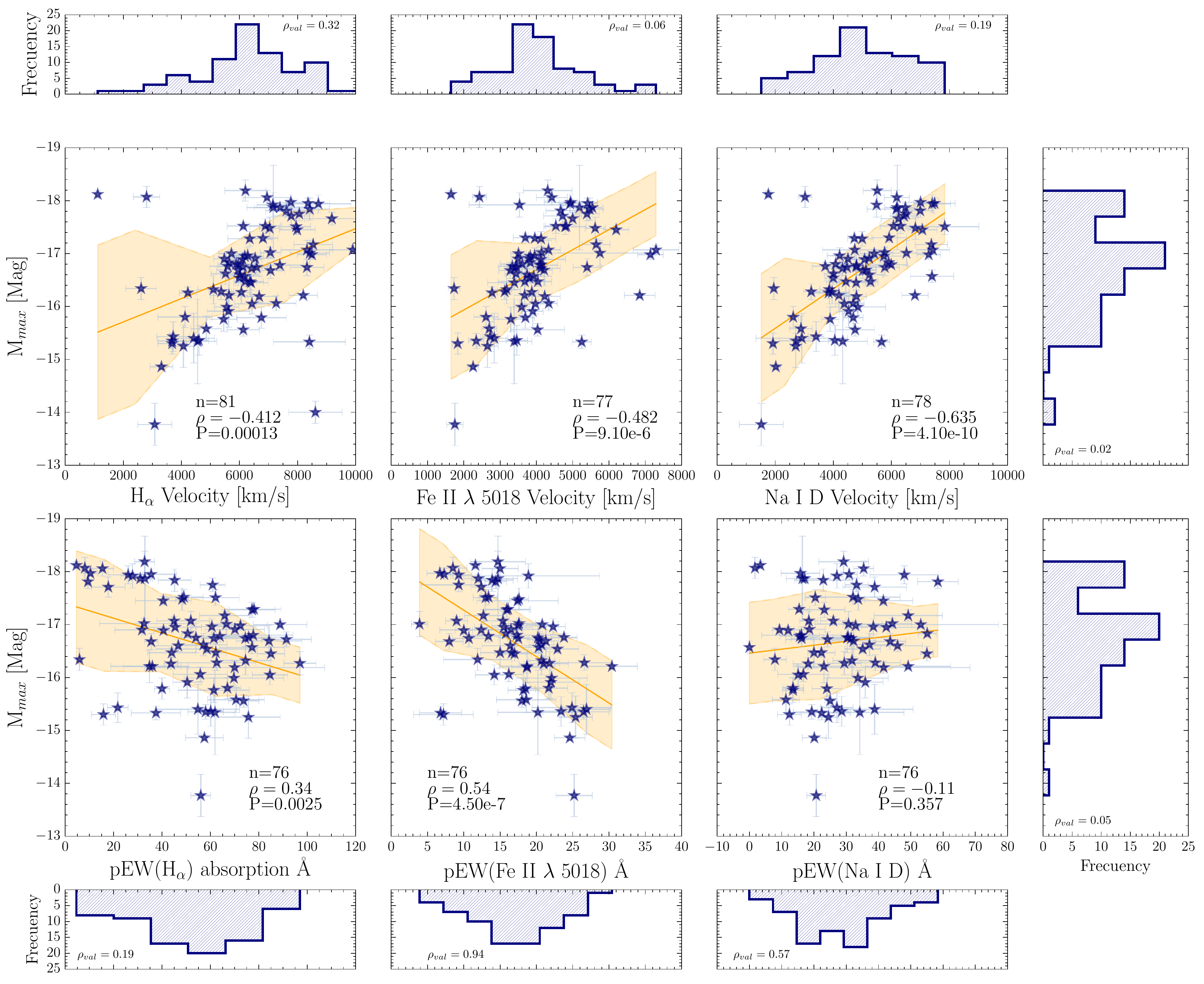}
\caption{\textbf{Top panel:} Correlations between 
M$_{max}$ and the expansion velocities. \textbf{Bottom panel:} Correlations between M$_{max}$ and the pEWs.
On the top of the figure: n = number of events, $\rho$ = Pearson’s
correlation coefficient, and P = probability of detecting a correlation by chance. In addition, each plot
shows the corresponding best fit (\texttt{linmix\_err}; \citealt{Kelly07}) as solid orange line, while the orange shaded
area indicates the variance with
respect to the fit line. Histograms along the x and y-axes show the distributions of the various parameters plotted on each axis.
Each histogram displays the $\rho_{val}$ found using the Shapiro-Wilk normalization. When the $\rho_{val}$ $>0.05$,
the dataset comes from a population which has a normal distribution.}
\label{mmaxvel}
\end{figure*}

\indent  Figure~\ref{mmaxvel} presents correlations between M$_{max}$ 
and the pEWs of H$_{\alpha}$, \ion{Fe}{2} 5018, and \ion{Na}{1} D. We observe a weak 
correlation with the pEW(H$_{\alpha}$) absorption component, a moderate ($\rho=0.54$) correlation
with pEW(\ion{Fe}{2} 5018), and no correlations with pEW(\ion{Na}{1} D). 

\begin{figure*}
\includegraphics[width=\textwidth]{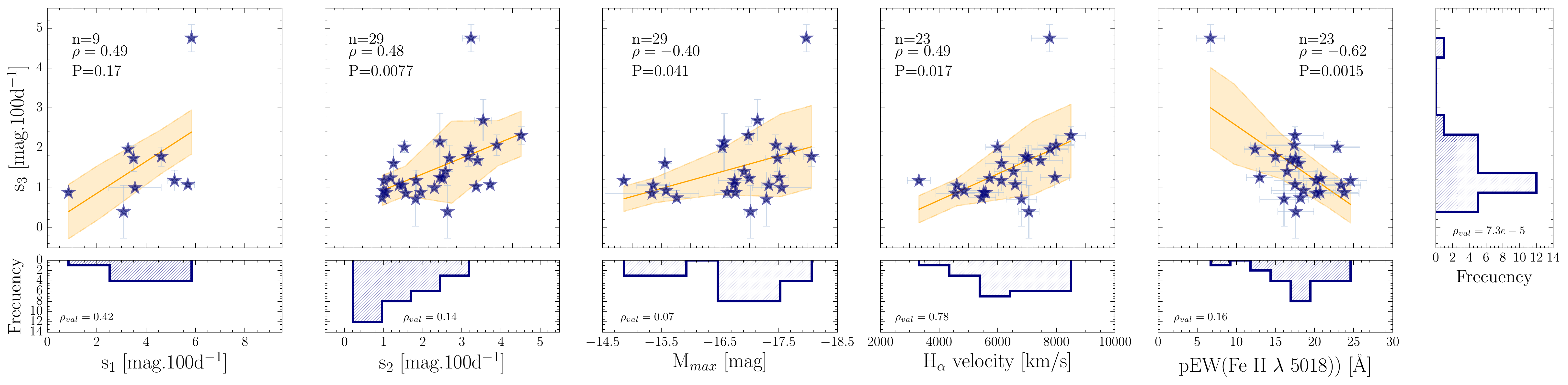}
\caption{Correlations between s$_3$ and five different parameters:
s$_1$, s$_2$, M$_{max}$, H$_{\alpha}$ velocity, pEW(\ion{Fe}{2} $\lambda$ 5018).
On the top of the figure: n = number of events, $\rho$ = Pearson’s
correlation coefficient, and P = probability of detecting a correlation by chance. Histograms
along the x and y-axes show the distributions of the various parameters plotted on each axis.
Each histogram displays the $\rho_{val}$ found using the Shapiro-Wilk normalization. When the $\rho_{val}$ $>0.05$,
the dataset comes from a population which has a normal distribution.}
\label{s3}
\end{figure*}

In Figure~\ref{s3} we repeat the correlations presented by A14, which show that a faster declining
SN at
one epoch is generally also a fast decliner at other epochs. 
Although the correlation of s$_3$ and M$_{max}$ is moderate, it is driven 
by an outlier event, SN~2006Y. As A14 noted, this SN presents an atypical 
behaviour in photometry, but here we confirm its strange behaviour in the spectra.
If we remove this SN from the analysis, the correlations decrease significantly. 
The correlations between s$_3$ and the velocities are moderate.
In the last panel of Figure~\ref{s3} the correlation between s$_3$
and the pEW(\ion{Fe}{2} 5018) is presented, which, like M$_{max}$ is driven by SN~2006Y. 
Summarizing, s$_3$ has weak correlations with the pEWs and the magnitudes.

\section{Discussion}
\label{discussion}

Using numerous defined spectral and photometric parameters we have searched for correlations 
between different observed properties of SNe II. We argue that $Pd$ is a better parameter than $OPTd$
for constraining the pre-SN hydrogen envelope mass.
Our analysis shows a strong correlation between $Pd$ and s$_3$, arguing that both of these parameters are
strongly linked to the hydrogen envelope mass/ejecta mass.
While expansion velocities and SN~II magnitudes display a significant degree of correlation, they show only
weak/moderate correlations with $Pd$ and s$_3$, suggesting
that explosion energy - observed through diversity in velocities and luminosity - and hydrogen envelope mass
vary somewhat independently between SNe~II.\\
\indent We now qualitatively compare our results with those found in previous studies, both observational and theoretical,
attempting to tie these correlations to the underlying physics of SNe~II. \\

\subsection{The influence of explosion energy}

\indent \citet{Hamuy02L} found that the luminosity of the SNe~IIP correlates with the photospheric velocity
(\ion{Fe}{2} velocity) at 50 days from explosion. Brighter SNe~II have higher ejecta expansion velocities.
This correlation has enabled the use of SNe~II as distance indicators.
In Figure~\ref{femag} we show the same relation, but in generalized form;
velocities correlate with SN~II brightness at all epochs. In addition, we show that this 
luminosity-velocity correlation is stronger at peak brightness (M$_{max}$) than during the plateau. 
\citet{Dessart13a} has shown that more energetic explosions produce more $^{56}$Ni mass, brighter SNe~II with faster
expanding velocities. 
This is consistent with our results, and suggests that explosion energy is indeed a 
primary parameter that influences SN~II diversity, and that is traced through SN~II brightness, velocities and $^{56}$Ni mass.

\subsection{The influence of hydrogen envelope mass}

\begin{figure}
\includegraphics[width=\columnwidth]{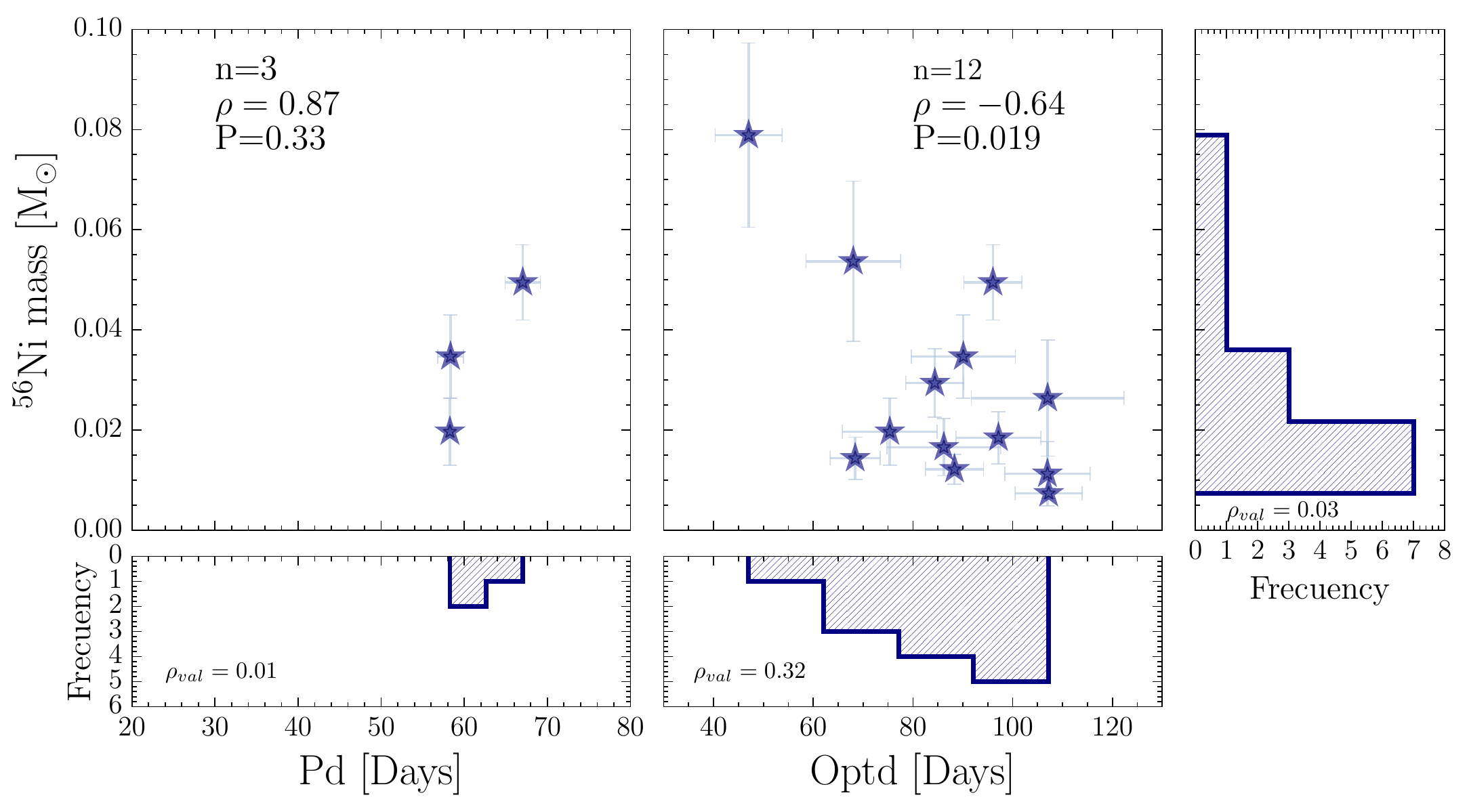}
\caption{Correlations between $^{56}$Ni mass and $Pd$ (on left) and $OPTd$ (on right).
On the top of the figure: n = number of events, $\rho$ = Pearson’s
correlation coefficient, and P = probability of detecting a correlation by chance.
Histograms along the x-and y-axes show the distributions of the various parameters plotted on each axis.
Each histogram displays the $\rho_{val}$ found using the Shapiro-Wilk normalization. When the $\rho_{val}$ $>0.05$,
the dataset comes from a population which has a normal distribution.}
\label{nitime}
\end{figure}

According to theoretical models faster declining SNe~II can be explained by the explosion of stars with
low hydrogen envelope mass (e.g. \citealt{Litvinova83,Bartunov92, Popov93} and \citealt{Moriya15}). 
As discussed previously, differences in envelope mass are likely to directly affect the length of the plateau, $Pd$ (we
again stress the difference between this parameter and $OPTd$, with the latter traditionally being assumed to be related to the
envelope mass). This is because the plateau, $Pd$, is powered by the recombination of hydrogen in the
expanding ejecta, and the lower the hydrogen envelope mass the quicker the recombination wave reaches its inner edge. The fact that $Pd$
also correlates with $s_3$ (Figure~\ref{pds3}) further supports this view, given that higher $s_3$ can be interpreted as being
due to a lower ejecta mass (A14) that can trap the radioactive emission (which is powering the light-curve at these late epochs).
With respect to faster declining SNe~II, we observe a significant trend in that SNe~II with higher $s_2$ have smaller $Pd$ values, implying
that the former is indeed related to the hydrogen envelope mass as has been predicted and discussed for many years.
Recent observational works \citep[e.g. A14,][]{Valenti16}
suggested that the phase between the explosion date and the end of the plateau (historically 
known as the plateau duration, but here named $OPTd$) is the key parameter constraining the envelope mass. 
However, $Pd$ shows higher degrees of correlation with other parameters, in particular $s_2$ and $s_3$. 
This suggests that $Pd$ is indeed a better tracer of envelope mass than $OPTd$. In addition, we find that a/e
shows strong and moderate correlation with $Pd$ and $s_3$ respectively, suggesting that this spectral parameter
is also a useful tracer of envelope mass (as already argued in \citealt{Gutierrez14}).\\
\indent From the the correlation matrix (Figure~\ref{pearson50}) we can observe stronger
relations between $Pd$ and $s_2$, as well as with 
the expansion velocity, than between $OPTd$ and the same parameters. This is because all these parameters are measured during the 
recombination phase, where they have similar physical conditions. On the other hand, $OPTd$ conveys information on the  
physical parameters that dominate the early phases of the light curve, plus the hydrogen envelope recombination. 
Consequently the correlations are weaker. \\
\indent In Figure~\ref{pearson50} we can see that $^{56}$Ni mass shows a strong correlation with $Pd$,
while with $OPTd$ display an anticorrelation. Analysing these findings (Figure~\ref{nitime}), we can see that
the relation between $^{56}$Ni mass and the $Pd$ is produced by only three measurements, and therefore the probability of this 
correlation being real is very small (P=0.33). In the case of the $OPTd$--$^{56}$Ni mass plot, this anti-correlation is
driven by a number of outliers.\\
\indent From Figure~\ref{pearson50}, we also see that $OPTd$ has stronger correlations with $Cd$, s$_1$ and M$_{tail}$ than
 with $Pd$. The strong relation between $OPTd$ and $Cd$ is expected because the former, by definition, 
includes the latter one (the same applies to $OPTd$ and $Pd$; see the $OPTd$ definition in Figure~\ref{lc}). However,
$Pd$ and $Cd$ are not related, because they are most likely associated with different physical properties of SNe~II. 
Between $OPTd$ and s$_1$ the correlation is moderate, but again, it is driven by the physical parameters that dominate
the early phases of the light curve, which, by definition, are included in $OPTd$.
One interesting correlation is displayed between $OPTd$ and M$_{tail}$: SNe~II with larger $OPTd$ values are fainter
in the radioactive tail phase. This relation may be understood given that the epoch of the M$_{tail}$ measurement 
directly arises from the length of $OPTd$. This means that, if the optically thick phase takes more time,
the M$_{tail}$ will be measured later, which in turn, implies fainter magnitudes (for the same $^{56}$Ni mass that 
is powering the late-time LC). This suggests that, the
correlation between $OPTd$ and M$_{tail}$ is essentially based on the total duration of the optically thick phase, i.e.,
the photospheric phase. \\
\indent In summary, we observe three key SN~II parameters that we believe are strongly related to the extent
of the hydrogen envelope mass at the moment of explosion: $Pd$, $s_3$, and a/e.
 
\subsection{The influence of explosion energy on the strength of spectral lines}

Figures~\ref{comp1},~\ref{comp2} and \ref{comp3} display some interesting trends. While the strength of each correlation is complicated by the obvious outliers
together with those SNe where no spectral line detection was made, in general it seems that expansion velocities correlate positively with the strength of the
Balmer lines and Na I D, and negatively with the strength of metal lines. The strength of metal lines at any given epoch is most strongly related to 
the temperature of the line forming region. We therefore conclude that more energetic explosions produce SNe~II that stay
at higher temperatures for longer leading to lower metal-line pEWs. With respect to the Balmer lines (at least the 
emission component of H$_{\alpha}$ and the absorption component of H$_{\beta}$) this would then imply that more energetic explosions lead to relatively
stronger line strengths. The exact physical interpretation of this is unclear. Brighter, i.e., more energetic SNe~II also display weaker
metal lines (Figure~\ref{pearson50} and specifically Figure~\ref{mmaxvel} bottom middle panel).
Finally, we also note that differences in progenitor metallicity 
will also affect the strength of metal lines within spectra, as argued by  \citet{Dessart14} and \citet{Anderson16}
(but probably to a lower degree, at least in the current sample).

\subsection{H$_{\alpha}$ P-Cygni diversity}
\label{halpha}

A large diversity in the H$_{\alpha}$ P-Cygni profile had been shown by \citet{Patat94} and \citet{Gutierrez14}.
They found that SNe~II with smaller $a/e$ values are brighter, and have higher velocities and steeper
decline rates. With our analysis at 50 days, we confirm these results, however
the correlations presented here are of lower strength than those in \citet{Gutierrez14}. 
This is most likely due to the epoch of the measurements, where in \citet{Gutierrez14} measurements were made 
at t$_{tran+10}$ (where t$_{tran}$ is the transitional epoch between s$_1$ and s$_2$).
Here we chose to use epochs with respect to explosion to measure our spectral parameters. This enables us 
to analyse the full range of events within our sample (in many SNe~II it is not possible to define t$_{tran}$).
The difference in correlation strength therefore arises from the measurements in \citet{Gutierrez14} being made when
SNe~II are likely to be under more consistent physical conditions. Here, using an epoch of 50 days post explosion different
SNe are at different phases of their evolution.\\
\indent It has previously been argued that the H$_{\alpha}$ P-Cygni diversity 
is directly related to the hydrogen envelope mass \citep{Schlegel96,Gutierrez14}. 
The results we present here also support this view, with the absorption component of H$_{\alpha}$ - and 
in particular the absorption in relation to the emission, a/e - showing 
correlation with both $Pd$ and s$_3$, parameters that we have already argued are direct tracers of the envelope
mass. We also note however that the measurement of H$_{\alpha}$ absorption is complicated by the detection and
diversity of Cachito (Paper I). It is quite possible therefore that vast majority of the underlying diversity
of H$_{\alpha}$ morphology is determined by the hydrogen envelope mass, but complications in the latter's
measurement introduce much of the dispersion we see (in e.g. Figure~\ref{pd}, bottom right).\\

\subsection{Other comparisons}

As discussed in \citet{Patat94}, A14 and more recently \citet{Valenti16} and \citet{Galbany16}, we find that 
faster declining SNe~II are brighter events (see Figure~\ref{pd}).  
In addition, we also find that SNe~II with brighter luminosities have greater expansion velocities and 
produce more $^{56}$Ni. In Figure~\ref{velni} and \ref{mmaxvel} we show a few examples of these correlations.
Similar results were found by several authors in observational \citep[e.g.][]{Hamuy03,Spiro14,Valenti16,Muller17}
and theoretical \citep[e.g.][]{Kasen09} works.\\
\indent Theoretical models show and increase in the $^{56}$Ni mass leads to an increase in the plateau duration 
(e.g.\citealt{Kasen09} and \citealt{Nakar16}). We do not find any observational evidence for such a trend. 
There are only 3 data points in the correlation between $Pd$ and $^{56}$Ni, therefore strong conclusions are not warranted.
If we include lower-mass $^{56}$Ni limits we also see no evidence for correlation. This may suggest that observationally 
$Pd$ does not depend on the mass of $^{56}$Ni mass. However, given the inclusion of lower-mass $^{56}$Ni 
limits, this warrants caution. \\
\indent Many authors have found \citep[e.g.][]{Dessart11} that SN~II color evolution could be related
with the radius of the progenitor star. Although we include the color gradient ($\Delta (B-V)$) between 
10--30 days post-explosion in our analysis, we do not find significant correlations associated to this parameter.
However, we do note low-level correlation between $\Delta (B-V)$ and the strength of \ion{Fe}{2} $\lambda$5018
and \ion{Fe}{2} $\lambda$5169 (Figure~\ref{pearson50}), in the direction one would expect: SNe~II that cool more
quickly (higher $\Delta (B-V)$) display stronger metal-line pEWs.
$Cd$ also does not display significant correlation with other parameters.
While above we linked $Cd$ to progenitor radius, as predicted by \citet[e.g.][]{Dessart13a},
the direct influence of radius on $Cd$ is complicated by any presence of CSM close to the progenitor
and may explain the lack of correlations.\\
\indent \citet{Dessart14} showed that differences in metallicity strongly influence in the SN~II spectra, more
precisely in the strength of the metal lines. \citet{Anderson16} supported this result showing a correlation between
the strength of \ion{Fe}{2} $\lambda$5018 with the oxygen abundance of host \ion{H}{2} regions. They showed
that SNe~II exploding in lower metallicity regions have lower iron absorption. Looking for relations with 
the pEW(\ion{Fe}{2} $\lambda$5018), we find a correlation of $0.48$ with the $Pd$ and $-0.62$ with s$_{3}$.
Assuming that the pEW(\ion{Fe}{2} $\lambda$5018) gives an idea of the metallicity where the SN explode, this correlation would mean that 
higher metallicity produce SNe with a longer plateau, which is in the opposite direction of the predictions
\citep[e.g.][]{Dessart13a}. However, when we correlate $Pd$ with the oxygen abundance
determined by \citet{Anderson16}, we do not find any relation. As in \citet{Anderson16}
we therefore conclude that (at least in the current sample), the strength of metal lines
is dependent more on temperature than progenitor metallicity.

\section{Conclusions}
\label{conclusions}

In this work we have presented an analysis of correlations between a range of spectral and photometric
parameters of 123 SNe~II, with the purpose of understanding their diversity. To study this diversity, we 
use the expansion velocities and pseudo-equivalent widths for eleven features in the 
photospheric phase (from explosion to $\sim120$ days): H$_{\alpha}$, H$_{\beta}$, \ion{Fe}{2} 4924,
\ion{Fe}{2} $\lambda$5018, \ion{Fe}{2} $\lambda$5169, \ion{Sc}{2}/\ion{Fe}{2} $\lambda$5531, \ion{Sc}{2} M, \ion{Na}{1} D, \ion{Ba}{2} $\lambda$6142, 
\ion{Sc}{2} $\lambda$6247, and  \ion{O}{1} $\lambda$7774; the ratio absorption to emission ($a/e$) of the H$_{\alpha}$ P-Cygni 
profile; the velocity decline rate of H$_{\beta}$ ($\Delta v$(H$_{\beta})$) and the velocity difference 
between H$_{\alpha}$ and Fe II $\lambda$5018, and Na I D and Fe II $\lambda$5018 ($\Delta vel$). From the light curves
we employed three magnitude measurements at different epochs (M$_{max}$, M$_{end}$, M$_{tail}$);
three decline rates (s$_1$, s$_2$, s$_3$); three time durations ($OPTd$, $Pd$, $Cd$); the  $^{56}$Ni mass,
and the color gradient, $\Delta (B-V)$. We searched for correlations at 30, 50 and 80 days, finding 
that correlations are stronger at 50 days post-explosion. We suggest this happens because 
at 50 days SNe~II are  under similar physical 
conditions: at 30 and 80 days not all SNe~II are in the same stage, some are in the cooling (at 
early phases) and some are in the transition to the nebular phase (at the end of the plateau).\\
\indent Our main results are summarized as follows:\\
\begin{itemize}
\item We confirm previous results showing that brighter SNe~II have higher expansion velocities. Here we
show that this finding is true for all SN~II decline rates, and also extends to magnitudes measured at maximum and during
the radioactive tail. These results are most easily explained through differences in explosion energy: more energetic
explosions produce brighter and higher velocity SNe~II. Additionally we find that more energetic (brighter and 
faster) events produce more $^{56}$Ni.
\item We highlight our different definition of the plateau duration ($Pd$) in this work as compared with the 
literature: from the s$_1$--s$_2$ transition to the end of the plateau, and conclude that it is a 
more robust parameter connected to H-rich envelope mass.
Indeed, we find that $Pd$ shows much stronger correlations with other parameters than
the traditionally used definition ($OPTd$ in our nomenclature). We conclude that $Pd$, s$_3$ and a/e are 
most directly affected by the hydrogen envelope mass at explosion epoch.
\item While we have found many different trends and correlations between different spectral and photometric 
parameters of SNe~II, hinting at underlying physical trends driving diversity (explosion energy, hydrogen 
envelope mass, $^{56}$Ni mass), we conclude there is no one parameter dominating these trends. 
\item As expected, expansion velocities measured for different spectral lines correlate strongly with each other.
However, velocities for different lines for individual SNe~II are significantly offset, suggesting that they form 
at different regions at differing distances from the photosphere.
\item Brighter SNe have higher velocities, smaller pEWs, shorter $a/e$, steeper declines and 
small $Pd$ and $OPTd$ values.

\end{itemize}

\acknowledgments
C.P.G., S.G.G. acknowledge support by projects IC120009 ``Millennium
Institute of Astrophysics (MAS)" and P10-064-F ``Millennium Center for Supernova
Science" of the Iniciativa Científica Milenio del Ministerio Economía,
Fomento y Turismo de Chile. C.P.G. acknowledges support from EU/FP7-ERC grant No. [615929].
M. D. S. is supported by the Danish Agency for Science and Technology and Innovation realized through a 
Sapere Aude Level 2 grant and by a research grant (13261) from the VILLUM FONDEN.
We gratefully acknowledge support of the CSP by the NSF under grants AST–0306969, AST–0908886, AST–0607438, and AST–1008343.
This research has made use of the NASA/IPAC
Extragalactic Database (NED) which is operated by the Jet Propulsion 
Laboratory, California Institute of Technology, under contract with the
National Aeronautics.

\clearpage
\LongTables

\clearpage
\begin{landscape}
\setcounter{table}{0}
\setlength\tabcolsep{2pt}
\tiny   
\begin{longtable}{lcccccccccccccccc} 
\caption{Photometric parameters}
\label{photo}\\
SN &  $Pd$ &  $OPTd$ &  $Cd$ &  M$_{max}$ &  M$_{end}$ &  M$_{tail}$ &  s$_{1}$            &  s$_{2}$ & s$_{3}$ &  $^{56}$Ni mass   &   $\Delta$(B-V)$_{10,30}$\\
    & (days)& (days)  & (days)& (mag)      & (mag)      & (mag)       &  (mag.100d$^{-1}$)  & (mag.100d$^{-1}$)  & (mag.100d$^{-1}$) & \Msun &  \\
\hline
\hline
1986L	& $59.56\pm0.71$  &  $93.74\pm6.71$     &  $34.18\pm3.08$     &    $-18.19\pm0.20$	&  $-16.88\pm0.20$  &  $-14.37\pm0.20$	&    $3.26\pm0.14$   &  $1.26\pm0.03$    &  \nodata	     &    $>$0.061                    &   $2.63\pm0.42$\\
1990K	& \nodata	  &  \nodata	        &  \nodata	      &    \nodata		&  \nodata	    &  \nodata		&    \nodata         &  $2.39\pm0.08$    &  \nodata	     &    \nodata                     &   \nodata	\\
1991al	& \nodata	  &  \nodata	        &  \nodata	      &    $-17.51\pm0.15$	&  $-17.03\pm0.15$  &  $-14.71\pm0.15$	&    \nodata         &  $1.45\pm0.04$    &  $1.26\pm0.26$    &    0.067$^{+0.016}_{-0.021}$   &   $3.69\pm0.24$\\
1992af	& \nodata	  &  $47.03\pm6.71$     &  \nodata	      &    $-17.33\pm0.12$	&  $-17.20\pm0.12$  &  $-15.06\pm0.12$	&    \nodata         &  $0.58\pm0.03$    &  $1.07\pm0.08$    &    0.079$^{+0.018}_{-0.029}$   &   $1.90\pm0.83$\\
\hline
\hline
\end{longtable}
\setcounter{table}{0}
\small
\begin{list}{}{}
\item Same as \citet{Anderson14}: In the first column we list the SN name. Columns 2, 3 and 4 shows the $Pd$, $OPTd$ and $Cd$. In columns 5, 6 and 7 we
list the absolute magnitudes of M$_{max}$, M$_{end}$ and M$_{tail}$ respectively. These are followed by the decline rates: s$_{1}$, s$_{2}$ and s$_{3}$, in columns 8, 9 and 
10 respectively. In column 11 we present the derived $^{56}$Ni masses (or lower limits), while in column 12 the color gradient is shown.
\item As it is explained in Section~\ref{measurements}, the $Pd$, s$_1$, s$2$ show differences with respect to \citet{Anderson14}. 
\end{list}

\clearpage
\end{landscape}

\clearpage
\setcounter{table}{1}
\begin{table*}[h!]
\small
\centering
\caption{Average of correlations}
\label{average}
\begin{tabular}{lcccccccccccccc}
Parameter & Average at 30 days & Average at 50 days & Average at 80 days\\
\hline
\hline
Pd	        			&   0.370    &	 0.410    &    0.425  \\
OPTd    				&   0.305    &   0.316    &    0.342  \\
Cd        				&   0.225    &   0.228    &    0.233  \\
M$_{max}$       			&   0.392    &   0.417    &    0.375  \\
M$_{end}$       			&   0.325    &   0.345    &    0.343  \\
M$_{tail}$      			&   0.406    &   0.423    &    0.456  \\
s$_1$        				&   0.355    &   0.391    &    0.344  \\
s$_2$        				&   0.304    &   0.348    &    0.325  \\
s$_3$        				&   0.334    &   0.374    &    0.363  \\
$^{56}$Ni        			&   0.449    &   0.520    &    0.550  \\
$\Delta C_({10-30})$      		&   0.208    &   0.219    &    0.213  \\
V(H$_{\alpha}$)        			&   0.361    &   0.468    &    0.452  \\
V(H$_{\beta}$)        			&   0.416    &   0.479    &    0.441  \\
V(Fe II 5018)        			&   0.380    &   0.450    &    0.325  \\
V(Fe II 5169)         			&   0.415    &   0.477    &    0.393  \\
V(Na I D)         			&   0.450    &   0.519    &    0.480  \\
pEW(H$_{\alpha})_a$   			&   0.279    &   0.270    &    0.287  \\
pEW(H$_{\alpha})_e$   	  		&   0.138    &   0.362    &    0.427  \\
pEW(Fe II 5018)       	 		&   0.329    &   0.339    &    0.218  \\
pEW(Fe II 5169)        			&   0.167    &   0.209    &    0.189  \\
pEW(Na I D)        			&   0.238    &   0.242    &    0.354  \\
$a/e$        				&   0.269    &   0.328    &    0.316  \\
$\Delta vel$(H$_{\alpha}$ - Fe II 5018) &   0.303    &   0.321    &    0.438  \\
$\Delta vel$(Na I D - Fe II 5018)       &   0.403    &   0.426    &    0.419  \\
$\Delta v$(H$_{\beta}$)  		&   0.248    &   0.228    &    0.207  \\
\hline
\end{tabular}
\begin{list}{}{}
\item Average of the correlations at 30, 50 and 80 days since explosion presented for 11 photometric parameters 
and 14 spectroscopic ones. In the first column the SN II parameter is listed (described in \ref{measurements}),
while in the second, three and four column are the average.
\end{list}
\end{table*} 

\clearpage
\begin{landscape}
\setcounter{table}{2}
\tiny
\begin{longtable}{lcccccccccccccccccc} 
\caption{Velocity values at 50 days from explosion}
\label{velocidades}\\
SN	&  $vel$(H$_{\alpha}$)	&  $vel$(H$_{\alpha}$)	&   $vel$(H$_{\beta}$)	&   $vel$(Fe 	&   $vel$(Fe II		&    $vel$(Fe II	&    $vel$(Fe II        &    $vel$(Sc II	&    $vel$(Na I D)	&    $vel$(Ba II)   	&    $vel$(ScII)  & $vel$(O I) \\[0.5ex]
	&			&			&			& II $\lambda$4924)& $\lambda$5018)  	& $\lambda$5169)        &     /Sc II)           &	 Mult.)		&			&			&		  &		\\[0.5ex]
	& (km s$^{-1}$)         & (km s$^{-1}$)         &  (km s$^{-1}$)        & (km s$^{-1}$) &  (km s$^{-1}$)        &  (km s$^{-1}$)        &  (km s$^{-1}$)        &  (km s$^{-1}$)        &  (km s$^{-1}$)        &  (km s$^{-1}$)        & (km s$^{-1}$)   &  (km s$^{-1}$) \\[0.5ex]
\hline
\hline
1986L	&  $7707\pm710$    &    $6204\pm476$	&    $6722\pm434$	&    $4946\pm245$	&    $4311\pm672$	&    $4406\pm411$	&      $4778\pm456$	&    $4316\pm336$	&    $5512\pm486$	&    \nodata        	&    \nodata         &   \nodata       \\[0.5ex]	
1990K	&  $7841\pm432$    &    $6958\pm426$	&    $6563\pm381$	&    \nodata		&    $4440\pm298$	&    $3942\pm204$	&      $4886\pm450$	&    $4604\pm239$	&    $6363\pm391$	&    $4468\pm312$	&    $4139\pm238$    &   \nodata       \\[0.5ex]	
1991al	&  $8897\pm496$    &    $7950\pm615$	&    $7789\pm634$	&    $4921\pm244$	&    $4790\pm588$	&    $4982\pm548$	&      $5256\pm261$	&    $4575\pm227$	&    $7835\pm876$	&    $5696\pm282$	&    $4149\pm206$    &   \nodata       \\
\hline
\hline
\end{longtable}
\setcounter{table}{0}
\small
\begin{list}{}{}
\item Columns: (1) SN name; (2) Velocity of H$_{\alpha}$ absorption component; (3) Velocity of H$_{\alpha}$ emission component; (4) Velocity of H$_{\beta}$; (5) Velocity of Fe II $\lambda$4924;
(6) Velocity of Fe II $\lambda$5018; (7) Velocity of Fe II $\lambda$5169; (8) Velocity of Fe II/Sc II; (9) Velocity of Sc II Multiplet; (10) Velocity of Na I D; (11) Velocity of Ba II; (12) Velocity of ScII;
and (13) Velocity of O I. 
\end{list}
\clearpage
\end{landscape}

\clearpage
\begin{landscape}
\setcounter{table}{3}
\tiny
\begin{longtable}{lcccccccccccccccc} 
\caption{pEW values at 50 days from explosion}
\label{pews}\\
SN	&  H$_{\alpha}$&  H$_{\alpha}$   &  H$_{\beta}$	& Fe II $\lambda$4924 & Fe II $\lambda$5018   & Fe II $\lambda$5169  & Fe II/Sc II  &  Sc II Mult. & Na I D       &  Ba II       &      ScII   & $a/e$\\[0.5ex]		
        &    (\AA)     &      (\AA)      &     (\AA)    &    (\AA)         &     (\AA)         &    (\AA)         &    (\AA)       &    (\AA)       &    (\AA)       &    (\AA)       &    (\AA)     &        \\[0.5ex]
\hline
\hline
1986L	&    $32.8\pm4.1$  &  $144.2\pm34.2$ &  $48.2\pm3.6$    &   $1.2\pm0.6$    &   $14.7\pm1.8$    &  $36.7\pm3.8$   &  $7.6\pm2.9$    &  $10.4\pm2.1$  &  $29.2\pm2.7$  &  $0.0\pm0.0$   &  $0.0\pm0.0$  &  $0.23\pm0.08$ \\[0.5ex]	
1988A	&    \nodata       &  \nodata        &  \nodata         &   \nodata        &   \nodata         &  \nodata         &  \nodata       &  \nodata       &  \nodata       &  \nodata       &  \nodata      &  \nodata \\[0.5ex]	
1990E	&    \nodata       &  \nodata        &  \nodata         &   \nodata        &   \nodata         &  \nodata         &  \nodata       &  \nodata       &  \nodata       &  \nodata       &  \nodata      &  \nodata \\[0.5ex]     
1990K	&    $42.7\pm3.8$  &  $206.2\pm27.4$ &  $71.9\pm4.4$    &   $0.0\pm0.0$    &   $10.9\pm0.7$    &  $38.8\pm2.1$    &  $8.9\pm0.7$   &  $13.2\pm1.6$  &  $50.3\pm2.5$  &  $6.4\pm0.4$   &  $5.9\pm0.9$  &  $0.21\pm0.05$ \\[0.5ex]   
1991al	&    $62.1\pm5.8$  &  $214.2\pm25.8$ &  $67.5\pm5.2$    &   $4.8\pm1.7$    &   $13.0\pm1.7$    &  $27.2\pm3.8$    &  $4.3\pm1.9$   &  $6.5\pm1.8$   &  $20.4\pm1.2$  &  $6.4\pm0.9$   &  $4.1\pm1.2$  &  $0.29\pm0.08$ \\
\hline
\hline
\end{longtable}
\setcounter{table}{0}
\small
\begin{list}{}{}
\item Columns: (1) SN name; (2) pEW of H$_{\alpha}$ absorption component; (3) pEW of H$_{\alpha}$ emission component; (4) pEW of H$_{\beta}$; (5) pEW of Fe II $\lambda$4924;
(6) pEW of Fe II $\lambda$5018; (7) pEW of Fe II $\lambda$5169; (8) pEW of Fe II/Sc II; (9) pEW of Sc II Multiplet; (10) pEW of Na I D; (11) pEW of Ba II; (12) pEW of ScII; 
(13) Ratio of absoprtion to emission ($a/e$) of H$_{\alpha}$ P-Cygni profile. \\	
\end{list}
\clearpage
\end{landscape}


\begin{thebibliography}{72}
\expandafter\ifx\csname natexlab\endcsname\relax\def\natexlab#1{#1}\fi

\bibitem[{{Anderson} {et~al.}(2014{\natexlab{a}}){Anderson}, {Dessart},
  {Gutierrez}, {Hamuy}, {Morrell}, {Phillips}, {Folatelli}, {Stritzinger},
  {Freedman}, {Gonz{\'a}lez-Gait{\'a}n}, {McCarthy}, {Suntzeff}, \&
  {Thomas-Osip}}]{Anderson14a}
{Anderson}, J.~P., {Dessart}, L., {Gutierrez}, C.~P., {et~al.}
  2014{\natexlab{a}}, \mnras, 441, 671

\bibitem[{{Anderson} {et~al.}(2014{\natexlab{b}})}]{Anderson14}
{Anderson}, J.~P., {et~al.} 2014{\natexlab{b}}, \apj, 786, 67

\bibitem[{{Anderson} {et~al.}(2016){Anderson}, {Guti{\'e}rrez}, {Dessart},
  {Hamuy}, {Galbany}, {Morrell}, {Stritzinger}, {Phillips}, {Folatelli},
  {Boffin}, {de Jaeger}, {Kuncarayakti}, \& {Prieto}}]{Anderson16}
{Anderson}, J.~P., {Guti{\'e}rrez}, C.~P., {Dessart}, L., {et~al.} 2016, \aap,
  589, A110

\bibitem[{{Barbon} {et~al.}(1979){Barbon}, {Ciatti}, \& {Rosino}}]{Barbon79}
{Barbon}, R., {Ciatti}, F., \& {Rosino}, L. 1979, \aap, 72, 287

\bibitem[{{Bartunov} \& {Blinnikov}(1992)}]{Bartunov92}
{Bartunov}, O.~S., \& {Blinnikov}, S.~I. 1992, Soviet Astronomy Letters, 18, 43

\bibitem[{{Bersten}(2013)}]{Bersten13T}
{Bersten}, M.~C. 2013, ArXiv e-prints

\bibitem[{{Blanco} {et~al.}(1987){Blanco}, {Gregory}, {Hamuy}, {Heathcote},
  {Phillips}, {Suntzeff}, {Terndrup}, {Walker}, {Williams}, {Pastoriza},
  {Storchi-Bergmann}, \& {Matthews}}]{Blanco87}
{Blanco}, V.~M., {Gregory}, B., {Hamuy}, M., {et~al.} 1987, \apj, 320, 589

\bibitem[{{Chevalier}(1976)}]{Chevalier76}
{Chevalier}, R.~A. 1976, \apj, 207, 872

\bibitem[{{Dessart} \& {Hillier}(2005)}]{Dessart05}
{Dessart}, L., \& {Hillier}, D.~J. 2005, \aap, 437, 667

\bibitem[{{Dessart} \& {Hillier}(2011)}]{Dessart11}
---. 2011, \mnras, 410, 1739

\bibitem[{{Dessart} {et~al.}(2017){Dessart}, {Hillier}, \& {Audit}}]{Dessart17}
{Dessart}, L., {Hillier}, D.~J., \& {Audit}, E. 2017, ArXiv e-prints

\bibitem[{{Dessart} {et~al.}(2013{\natexlab{a}}){Dessart}, {Hillier},
  {Waldman}, \& {Livne}}]{Dessart13a}
{Dessart}, L., {Hillier}, D.~J., {Waldman}, R., \& {Livne}, E.
  2013{\natexlab{a}}, \mnras, 433, 1745

\bibitem[{{Dessart} {et~al.}(2010{\natexlab{a}}){Dessart}, {Livne}, \&
  {Waldman}}]{Dessart10b}
{Dessart}, L., {Livne}, E., \& {Waldman}, R. 2010{\natexlab{a}}, \mnras, 408,
  827

\bibitem[{{Dessart} {et~al.}(2010{\natexlab{b}}){Dessart}, {Livne}, \&
  {Waldman}}]{Dessart10}
---. 2010{\natexlab{b}}, \mnras, 405, 2113

\bibitem[{{Dessart} {et~al.}(2013{\natexlab{b}}){Dessart}, {Waldman}, {Livne},
  {Hillier}, \& {Blondin}}]{Dessart13}
{Dessart}, L., {Waldman}, R., {Livne}, E., {Hillier}, D.~J., \& {Blondin}, S.
  2013{\natexlab{b}}, \mnras, 428, 3227

\bibitem[{{Dessart} {et~al.}(2014){Dessart}, {Gutierrez}, {Hamuy}, {Hillier},
  {Lanz}, {Anderson}, {Folatelli}, {Freedman}, {Ley}, {Morrell}, {Persson},
  {Phillips}, {Stritzinger}, \& {Suntzeff}}]{Dessart14}
{Dessart}, L., {Gutierrez}, C.~P., {Hamuy}, M., {et~al.} 2014, \mnras, 440,
  1856

\bibitem[{{Elias-Rosa} {et~al.}(2010){Elias-Rosa}, {Van Dyk}, {Li}, {Miller},
  {Silverman}, {Ganeshalingam}, {Boden}, {Kasliwal}, {Vink{\'o}}, {Cuillandre},
  {Filippenko}, {Steele}, {Bloom}, {Griffith}, {Kleiser}, \&
  {Foley}}]{Elias-Rosa10}
{Elias-Rosa}, N., {Van Dyk}, S.~D., {Li}, W., {et~al.} 2010, \apjl, 714, L254

\bibitem[{{Elias-Rosa} {et~al.}(2011){Elias-Rosa}, {Van Dyk}, {Li},
  {Silverman}, {Foley}, {Ganeshalingam}, {Mauerhan}, {Kankare}, {Jha},
  {Filippenko}, {Beckman}, {Berger}, {Cuillandre}, \& {Smith}}]{Elias-Rosa11}
---. 2011, \apj, 742, 6

\bibitem[{{Evans}(1996)}]{Evans96}
{Evans}, J.~D. 1996, Pacific Grove, CA: Brooks/Cole Publishing

\bibitem[{{Falk} \& {Arnett}(1977)}]{Falk77}
{Falk}, S.~W., \& {Arnett}, W.~D. 1977, \apjs, 33, 515

\bibitem[{{Faran} {et~al.}(2014{\natexlab{a}}){Faran}, {Poznanski},
  {Filippenko}, {Chornock}, {Foley}, {Ganeshalingam}, {Leonard}, {Li},
  {Modjaz}, {Serduke}, \& {Silverman}}]{Faran14b}
{Faran}, T., {Poznanski}, D., {Filippenko}, A.~V., {et~al.} 2014{\natexlab{a}},
  \mnras, 445, 554

\bibitem[{{Faran} {et~al.}(2014{\natexlab{b}}){Faran}, {Poznanski},
  {Filippenko}, {Chornock}, {Foley}, {Ganeshalingam}, {Leonard}, {Li},
  {Modjaz}, {Nakar}, {Serduke}, \& {Silverman}}]{Faran14a}
---. 2014{\natexlab{b}}, \mnras, 442, 844

\bibitem[{{Filippenko}(1997)}]{Filippenko97}
{Filippenko}, A.~V. 1997, \araa, 35, 309

\bibitem[{{Filippenko} {et~al.}(1993){Filippenko}, {Matheson}, \&
  {Ho}}]{Filippenko93}
{Filippenko}, A.~V., {Matheson}, T., \& {Ho}, L.~C. 1993, \apjl, 415, L103

\bibitem[{{Galbany} {et~al.}(2016){Galbany}, {Hamuy}, {Phillips}, {Suntzeff},
  {Maza}, {de Jaeger}, {Moraga}, {Gonz{\'a}lez-Gait{\'a}n}, {Krisciunas},
  {Morrell}, {Thomas-Osip}, {Krzeminski}, {Gonz{\'a}lez}, {Antezana},
  {Wishnjewski}, {McCarthy}, {Anderson}, {Guti{\'e}rrez}, {Stritzinger},
  {Folatelli}, {Anguita}, {Galaz}, {Green}, {Impey}, {Kim}, {Kirhakos},
  {Malkan}, {Mulchaey}, {Phillips}, {Pizzella}, {Prosser}, {Schmidt},
  {Schommer}, {Sherry}, {Strolger}, {Wells}, \& {Williger}}]{Galbany16}
{Galbany}, L., {Hamuy}, M., {Phillips}, M.~M., {et~al.} 2016, \aj, 151, 33

\bibitem[{{Gonz{\'a}lez-Gait{\'a}n} {et~al.}(2015){Gonz{\'a}lez-Gait{\'a}n},
  {Tominaga}, {Molina}, {Galbany}, {Bufano}, {Anderson}, {Gutierrez},
  {F{\"o}rster}, {Pignata}, {Bersten}, {Howell}, {Sullivan}, {Carlberg}, {de
  Jaeger}, {Hamuy}, {Baklanov}, \& {Blinnikov}}]{Gonzalez15}
{Gonz{\'a}lez-Gait{\'a}n}, S., {Tominaga}, N., {Molina}, J., {et~al.} 2015,
  \mnras, 451, 2212

\bibitem[{{Grassberg} {et~al.}(1971){Grassberg}, {Imshennik}, \&
  {Nadyozhin}}]{Grassberg71}
{Grassberg}, E.~K., {Imshennik}, V.~S., \& {Nadyozhin}, D.~K. 1971, \apss, 10,
  28

\bibitem[{{Guti{\'e}rrez} {et~al.}(2014)}]{Gutierrez14}
{Guti{\'e}rrez}, C.~P., {et~al.} 2014, \apjl, 786, L15

\bibitem[{{Hamuy}(2003)}]{Hamuy03}
{Hamuy}, M. 2003, \apj, 582, 905

\bibitem[{{Hamuy} \& {Pinto}(2002)}]{Hamuy02L}
{Hamuy}, M., \& {Pinto}, P.~A. 2002, \apjl, 566, L63

\bibitem[{{Hamuy} {et~al.}(1988){Hamuy}, {Suntzeff}, {Gonzalez}, \&
  {Martin}}]{Hamuy88}
{Hamuy}, M., {Suntzeff}, N.~B., {Gonzalez}, R., \& {Martin}, G. 1988, \aj, 95,
  63

\bibitem[{{Hamuy} {et~al.}(2002){Hamuy}, {Maza}, {Pinto}, {Phillips},
  {Suntzeff}, {Blum}, {Olsen}, {Pinfield}, {Ivanov}, {Augusteijn}, {Brillant},
  {Chadid}, {Cuby}, {Doublier}, {Hainaut}, {Le Floc'h}, {Lidman},
  {Petr-Gotzens}, {Pompei}, \& {Vanzi}}]{Hamuy02}
{Hamuy}, M., {Maza}, J., {Pinto}, P.~A., {et~al.} 2002, \aj, 124, 417

\bibitem[{{Inserra} {et~al.}(2013){Inserra}, {Pastorello}, {Turatto}, {Pumo},
  {Benetti}, {Cappellaro}, {Botticella}, {Bufano}, {Elias-Rosa}, {Harutyunyan},
  {Taubenberger}, {Valenti}, \& {Zampieri}}]{Inserra13}
{Inserra}, C., {Pastorello}, A., {Turatto}, M., {et~al.} 2013, \aap, 555, A142

\bibitem[{{Kasen} \& {Woosley}(2009)}]{Kasen09}
{Kasen}, D., \& {Woosley}, S.~E. 2009, \apj, 703, 2205

\bibitem[{{Kelly}(2007)}]{Kelly07}
{Kelly}, B.~C. 2007, \apj, 665, 1489

\bibitem[{{Litvinova} \& {Nadezhin}(1983)}]{Litvinova83}
{Litvinova}, I.~I., \& {Nadezhin}, D.~K. 1983, \apss, 89, 89

\bibitem[{{Maund} {et~al.}(2015){Maund}, {Fraser}, {Reilly}, {Ergon}, \&
  {Mattila}}]{Maund15}
{Maund}, J.~R., {Fraser}, M., {Reilly}, E., {Ergon}, M., \& {Mattila}, S. 2015,
  \mnras, 447, 3207

\bibitem[{{Maund} \& {Smartt}(2005)}]{Maund05}
{Maund}, J.~R., \& {Smartt}, S.~J. 2005, \mnras, 360, 288

\bibitem[{{Menzies} {et~al.}(1987){Menzies}, {Catchpole}, {van Vuuren},
  {Winkler}, {Laney}, {Whitelock}, {Cousins}, {Carter}, {Marang}, {Lloyd
  Evans}, {Roberts}, {Kilkenny}, {Spencer Jones}, {Sekiguchi}, {Fairall}, \&
  {Wolstencroft}}]{Menzies87}
{Menzies}, J.~W., {Catchpole}, R.~M., {van Vuuren}, G., {et~al.} 1987, \mnras,
  227, 39P

\bibitem[{{Minkowski}(1941)}]{Minkowski41}
{Minkowski}, R. 1941, \pasp, 53, 224

\bibitem[{{Moriya} {et~al.}(2016){Moriya}, {Pruzhinskaya}, {Ergon}, \&
  {Blinnikov}}]{Moriya15}
{Moriya}, T.~J., {Pruzhinskaya}, M.~V., {Ergon}, M., \& {Blinnikov}, S.~I.
  2016, \mnras, 455, 423

\bibitem[{{Moriya} {et~al.}(2017){Moriya}, {Yoon}, {Gr{\"a}fener}, \&
  {Blinnikov}}]{Moriya17}
{Moriya}, T.~J., {Yoon}, S.-C., {Gr{\"a}fener}, G., \& {Blinnikov}, S.~I. 2017,
  \mnras, 469, L108

\bibitem[{{Morozova} {et~al.}(2016){Morozova}, {Piro}, {Renzo}, \&
  {Ott}}]{Morozova16}
{Morozova}, V., {Piro}, A.~L., {Renzo}, M., \& {Ott}, C.~D. 2016, \apj, 829,
  109

\bibitem[{{Morozova} {et~al.}(2015){Morozova}, {Piro}, {Renzo}, {Ott},
  {Clausen}, {Couch}, {Ellis}, \& {Roberts}}]{Morozova15}
{Morozova}, V., {Piro}, A.~L., {Renzo}, M., {et~al.} 2015, \apj, 814, 63

\bibitem[{{Morozova} {et~al.}(2017){Morozova}, {Piro}, \&
  {Valenti}}]{Morozova17}
{Morozova}, V., {Piro}, A.~L., \& {Valenti}, S. 2017, \apj, 838, 28

\bibitem[{{M{\"u}ller} {et~al.}(2017){M{\"u}ller}, {Prieto}, {Pejcha}, \&
  {Clocchiatti}}]{Muller17}
{M{\"u}ller}, T., {Prieto}, J.~L., {Pejcha}, O., \& {Clocchiatti}, A. 2017,
  \apj, 841, 127

\bibitem[{{Nakar} {et~al.}(2016){Nakar}, {Poznanski}, \& {Katz}}]{Nakar16}
{Nakar}, E., {Poznanski}, D., \& {Katz}, B. 2016, \apj, 823, 127

\bibitem[{{Pastorello} {et~al.}(2003){Pastorello}, {Ramina}, {Zampieri},
  {Navasardyan}, {Salvo}, \& {Fiaschi}}]{Pastorello03}
{Pastorello}, A., {Ramina}, M., {Zampieri}, L., {et~al.} 2003, ArXiv
  Astrophysics e-prints

\bibitem[{{Pastorello} {et~al.}(2004){Pastorello}, {Zampieri}, {Turatto},
  {Cappellaro}, {Meikle}, {Benetti}, {Branch}, {Baron}, {Patat}, {Armstrong},
  {Altavilla}, {Salvo}, \& {Riello}}]{Pastorello04}
{Pastorello}, A., {Zampieri}, L., {Turatto}, M., {et~al.} 2004, \mnras, 347, 74

\bibitem[{{Pastorello} {et~al.}(2005){Pastorello}, {Baron}, {Branch},
  {Zampieri}, {Turatto}, {Ramina}, {Benetti}, {Cappellaro}, {Salvo}, {Patat},
  {Piemonte}, {Sollerman}, {Leibundgut}, \& {Altavilla}}]{Pastorello05}
{Pastorello}, A., {Baron}, E., {Branch}, D., {et~al.} 2005, \mnras, 360, 950

\bibitem[{{Pastorello} {et~al.}(2012){Pastorello}, {Pumo}, {Navasardyan},
  {Zampieri}, {Turatto}, {Sollerman}, {Taddia}, {Kankare}, {Mattila},
  {Nicolas}, {Prosperi}, {San Segundo Delgado}, {Taubenberger}, {Boles},
  {Bachini}, {Benetti}, {Bufano}, {Cappellaro}, {Cason}, {Cetrulo}, {Ergon},
  {Germany}, {Harutyunyan}, {Howerton}, {Hurst}, {Patat}, {Stritzinger},
  {Strolger}, \& {Wells}}]{Pastorello12}
{Pastorello}, A., {Pumo}, M.~L., {Navasardyan}, H., {et~al.} 2012, \aap, 537,
  A141

\bibitem[{{Patat} {et~al.}(1994){Patat}, {Barbon}, {Cappellaro}, \&
  {Turatto}}]{Patat94}
{Patat}, F., {Barbon}, R., {Cappellaro}, E., \& {Turatto}, M. 1994, \aap, 282,
  731

\bibitem[{{Pejcha} \& {Prieto}(2015{\natexlab{a}})}]{Pejcha15}
{Pejcha}, O., \& {Prieto}, J.~L. 2015{\natexlab{a}}, \apj, 799, 215

\bibitem[{{Pejcha} \& {Prieto}(2015{\natexlab{b}})}]{Pejcha15a}
---. 2015{\natexlab{b}}, \apj, 806, 225

\bibitem[{{Phillips} {et~al.}(1988){Phillips}, {Heathcote}, {Hamuy}, \&
  {Navarrete}}]{Phillips88}
{Phillips}, M.~M., {Heathcote}, S.~R., {Hamuy}, M., \& {Navarrete}, M. 1988,
  \aj, 95, 1087

\bibitem[{{Popov}(1993)}]{Popov93}
{Popov}, D.~V. 1993, \apj, 414, 712

\bibitem[{{Roy} {et~al.}(2011){Roy}, {Kumar}, {Benetti}, {Pastorello}, {Yuan},
  {Brown}, {Immler}, {Fatkhullin}, {Moskvitin}, {Maund}, {Akerlof}, {Wheeler},
  {Sokolov}, {Quimby}, {Bufano}, {Kumar}, {Misra}, {Pandey}, {Elias-Rosa},
  {Roming}, \& {Sagar}}]{Roy11}
{Roy}, R., {Kumar}, B., {Benetti}, S., {et~al.} 2011, \apj, 736, 76

\bibitem[{{Sanders} {et~al.}(2015){Sanders}, {Soderberg}, {Gezari},
  {Betancourt}, {Chornock}, {Berger}, {Foley}, {Challis}, {Drout}, {Kirshner},
  {Lunnan}, {Marion}, {Margutti}, {McKinnon}, {Milisavljevic}, {Narayan},
  {Rest}, {Kankare}, {Mattila}, {Smartt}, {Huber}, {Burgett}, {Draper},
  {Hodapp}, {Kaiser}, {Kudritzki}, {Magnier}, {Metcalfe}, {Morgan}, {Price},
  {Tonry}, {Wainscoat}, \& {Waters}}]{Sanders14}
{Sanders}, N.~E., {Soderberg}, A.~M., {Gezari}, S., {et~al.} 2015, \apj, 799,
  208

\bibitem[{{Schlegel}(1990)}]{Schlegel90}
{Schlegel}, E.~M. 1990, \mnras, 244, 269

\bibitem[{{Schlegel}(1996)}]{Schlegel96}
---. 1996, \aj, 111, 1660

\bibitem[{{Smartt}(2015)}]{Smartt15}
{Smartt}, S.~J. 2015, Publications of the Astronomical Society of Australia, 32, e016

\bibitem[{{Smartt} {et~al.}(2009){Smartt}, {Eldridge}, {Crockett}, \&
  {Maund}}]{Smartt09}
{Smartt}, S.~J., {Eldridge}, J.~J., {Crockett}, R.~M., \& {Maund}, J.~R. 2009,
  \mnras, 395, 1409

\bibitem[{{Smartt} {et~al.}(2004){Smartt}, {Maund}, {Hendry}, {Tout},
  {Gilmore}, {Mattila}, \& {Benn}}]{Smartt04}
{Smartt}, S.~J., {Maund}, J.~R., {Hendry}, M.~A., {et~al.} 2004, Science, 303,
  499

\bibitem[{{Spiro} {et~al.}(2014){Spiro}, {Pastorello}, {Pumo}, {Zampieri},
  {Turatto}, {Smartt}, {Benetti}, {Cappellaro}, {Valenti}, {Agnoletto},
  {Altavilla}, {Aoki}, {Brocato}, {Corsini}, {Di Cianno}, {Elias-Rosa},
  {Hamuy}, {Enya}, {Fiaschi}, {Folatelli}, {Desidera}, {Harutyunyan}, {Howell},
  {Kawka}, {Kobayashi}, {Leibundgut}, {Minezaki}, {Navasardyan}, {Nomoto},
  {Mattila}, {Pietrinferni}, {Pignata}, {Raimondo}, {Salvo}, {Schmidt},
  {Sollerman}, {Spyromilio}, {Taubenberger}, {Valentini}, {Vennes}, \&
  {Yoshii}}]{Spiro14}
{Spiro}, S., {Pastorello}, A., {Pumo}, M.~L., {et~al.} 2014, \mnras, 439, 2873

\bibitem[{{Suntzeff} {et~al.}(1988){Suntzeff}, {Hamuy}, {Martin}, {Gomez}, \&
  {Gonzalez}}]{Suntzeff88}
{Suntzeff}, N.~B., {Hamuy}, M., {Martin}, G., {Gomez}, A., \& {Gonzalez}, R.
  1988, \aj, 96, 1864

\bibitem[{{Taddia} {et~al.}(2013)}]{Taddia13}
{Taddia}, F., {et~al.} 2013, \aap, 555, A10

\bibitem[{{Tak{\'a}ts} {et~al.}(2014){Tak{\'a}ts}, {Pumo}, {Elias-Rosa},
  {Pastorello}, {Pignata}, {Paillas}, {Zampieri}, {Anderson}, {Vink{\'o}},
  {Benetti}, {Botticella}, {Bufano}, {Campillay}, {Cartier}, {Ergon},
  {Folatelli}, {Foley}, {F{\"o}rster}, {Hamuy}, {Hentunen}, {Kankare},
  {Leloudas}, {Morrell}, {Nissinen}, {Phillips}, {Smartt}, {Stritzinger},
  {Taubenberger}, {Valenti}, {Van Dyk}, {Haislip}, {LaCluyze}, {Moore}, \&
  {Reichart}}]{Takats14}
{Tak{\'a}ts}, K., {Pumo}, M.~L., {Elias-Rosa}, N., {et~al.} 2014, \mnras, 438,
  368

\bibitem[{{Valenti} {et~al.}(2016){Valenti}, {Howell}, {Stritzinger}, {Graham},
  {Hosseinzadeh}, {Arcavi}, {Bildsten}, {Jerkstrand}, {McCully}, {Pastorello},
  {Piro}, {Sand}, {Smartt}, {Terreran}, {Baltay}, {Benetti}, {Brown},
  {Filippenko}, {Fraser}, {Rabinowitz}, {Sullivan}, \& {Yuan}}]{Valenti16}
{Valenti}, S., {Howell}, D.~A., {Stritzinger}, M.~D., {et~al.} 2016, \mnras,
  459, 3939

\bibitem[{{Van Dyk} {et~al.}(2003){Van Dyk}, {Li}, \& {Filippenko}}]{VanDyk03}
{Van Dyk}, S.~D., {Li}, W., \& {Filippenko}, A.~V. 2003, \pasp, 115, 1289

\bibitem[{{Woosley} {et~al.}(1989){Woosley}, {Hartmann}, \&
  {Pinto}}]{Woosley89}
{Woosley}, S.~E., {Hartmann}, D., \& {Pinto}, P.~A. 1989, \apj, 346, 395

\bibitem[{{Yaron} {et~al.}(2017){Yaron}, {Perley}, {Gal-Yam}, {Groh}, {Horesh},
  {Ofek}, {Kulkarni}, {Sollerman}, {Fransson}, {Rubin}, {Szabo}, {Sapir},
  {Taddia}, {Cenko}, {Valenti}, {Arcavi}, {Howell}, {Kasliwal}, {Vreeswijk},
  {Khazov}, {Fox}, {Cao}, {Gnat}, {Kelly}, {Nugent}, {Filippenko}, {Laher},
  {Wozniak}, {Lee}, {Rebbapragada}, {Maguire}, {Sullivan}, \&
  {Soumagnac}}]{Yaron17}
{Yaron}, O., {Perley}, D.~A., {Gal-Yam}, A., {et~al.} 2017, Nature Physics, 13,
  510

\bibitem[{{Young}(2004)}]{Young04}
{Young}, T.~R. 2004, \apj, 617, 1233

\end{thebibliography}
\end{document}